\documentclass[conference]{IEEEtran}

\usepackage{tikz}
\usepackage{amsmath}
\usepackage{filecontents}

\usepackage{silence}
\usepackage[hyphens]{url}
\usepackage[breaklinks,colorlinks]{hyperref}

\usepackage{breakurl}
\hypersetup{citecolor=blue,linkcolor=blue}
\usepackage{amsfonts}

\usepackage{caption}
\usepackage{subcaption}

\usepackage{endnotes}
\usepackage{microtype}
\usepackage{graphicx}
\usepackage{fancyvrb}
\usepackage{multirow}
\usepackage{booktabs}
\usepackage{array,underscore,relsize}
\usepackage{fancyhdr}
\let\labelindent\relax
\usepackage{enumitem}
\usepackage[labelfont=bf,font=small,skip=5pt]{caption}
\usepackage{xspace}
\usepackage{fp}
\usepackage{siunitx}

\usepackage{balance}

\usepackage{pifont}
\usepackage{wasysym}
\usepackage{cite}
\usepackage{algorithmic}
\usepackage{tikz}
\usepackage{textcomp}
\usepackage{bm}
\usepackage[10pt]{moresize}
\usepackage{colortbl}

\makeatletter
\renewcommand\footnoterule{%
  \kern-3\p@
  \hrule\@width\columnwidth
  \kern2.6\p@}
  \makeatother

\newcommand{\sys}{\mbox{\textsc{Mardu}}\xspace}

\newcommand{\cc}[1]{\mbox{\smaller[0.5]\texttt{#1}}}

\fvset{fontsize=\scriptsize,xleftmargin=8pt,numbers=left,numbersep=5pt}

\input{code/fmt}

\if 0
\fi

\def\Snospace~{\S{}}

\if 0

\fi

\newif\ifdraft\drafttrue
\newif\ifnotes\notestrue
\ifdraft\else\notesfalse\fi

\input{glyphtounicode}
\newcolumntype{R}[1]{>{\raggedleft\let\newline\\\arraybackslash\hspace{0pt}}p{#1}}

\newcommand{\includepdf}[1]{
  \includegraphics[width=\columnwidth]{#1}
}

\newcommand{\squishlist}{
\begin{itemize}[noitemsep,nolistsep]
  \setlength{\itemsep}{-0pt}
}
\newcommand{\squishend}{
  \end{itemize}
}

\newcommand{\squishlists}{
\begin{itemize}[leftmargin=3mm,noitemsep,nolistsep]
  \setlength{\itemsep}{-0pt}
}
\newcommand{\squishends}{
  \end{itemize}
}

\newcommand{\nullsquishlist}{
\begin{description}[noitemsep,nolistsep]
  \setlength{\itemsep}{-0pt}
}
\newcommand{\nullsquishend}{
  \end{description}
}
\usepackage{tikz}

\newcommand*\BC[1]{%
\begin{tikzpicture}[baseline=(C.base)]
\node[draw,circle,fill=black,inner sep=0.2pt](C) {\textcolor{white}{#1}};
\end{tikzpicture}}

\usepackage{xstring}
\newcommand{\PP}[1]{
\vspace{2px}
\noindent{\bf \IfEndWith{#1}{.}{#1}{#1.}}
}

\newcommand{\ra}[1]{\renewcommand{\arraystretch}{#1}}

\newcommand{\etal}{\emph{et al.}\xspace}
\newcommand{\etc}{\emph{etc.}\xspace}
\newcommand{\ie}{\emph{i.e.}}
\newcommand{\eg}{\emph{e.g.}}

\newcommand{\boxbeg}{
\vspace{5px}
\noindent
\begin{tabular}{|l|}\hline
\begin{minipage}{3.2in}
\vspace{3px}
\noindent
}

\newcommand{\boxend}{
\vspace{3px}
\end{minipage}\\ \hline
\end{tabular}
\vspace{5px}
}

\newcommand{\densetbl}{\addtolength{\tabcolsep}{-3pt}}
\newcommand{\densetblend}{\addtolength{\tabcolsep}{3pt}}

\newcommand{\rip}{\cc{\%rip}\xspace}

\newcommand{\rbp}{\cc{\%rbp}\xspace}

\newcommand{\gs}{\cc{\%gs}\xspace}

\newcommand{\jmp}{\cc{jmp}\xspace}
\newcommand{\call}{\cc{call}\xspace}
\newcommand{\ret}{\cc{ret}\xspace}

\newcommand{\rdrand}{\cc{rdrand}\xspace}
\newcommand{\rdseed}{\cc{rdseed}\xspace}

\newcommand{\fork}{\cc{fork}\xspace}
\newcommand{\clone}{\cc{clone}\xspace}
\newcommand{\mmap}{\cc{mmap}\xspace}

\newcommand{\libc}{\cc{libc.so}\xspace}

\newcommand{\sigbus}{\cc{SIGBUS}\xspace}

\newcommand{\inode}{\cc{inode}\xspace}

\newcommand{\fnfoo}{\cc{foo()}\xspace}
\newcommand{\fnbar}{\cc{bar()}\xspace}

\newcommand{\machinefunctionpass}{\cc{MachineFunctionPass}\xspace}

\newcommand\llvmAdded{3530}
\newcommand\llvmDeleted{19}
\newcommand\linuxAdded{3998}
\newcommand\linuxDeleted{11}
\newcommand\muslAdded{152}
\newcommand\muslDeleted{12}
\newcommand\totalAdded{\the\numexpr\llvmAdded+\linuxAdded\relax}
\newcommand\totalDeleted{\the\numexpr\llvmDeleted+\linuxDeleted\relax}
\newcommand\totalModified{\the\numexpr\totalAdded+\totalDeleted\relax}

\newcommand\totalModifiedLLVM{\the\numexpr\llvmAdded+\llvmDeleted\relax}
\newcommand\totalModifiedKernel{\the\numexpr\linuxAdded+\linuxDeleted\relax}
\newcommand\totalModifiedMusl{\the\numexpr\muslAdded+\muslDeleted\relax}
\newcommand\RETperlbench{39174}
\newcommand\RETbzip{926}
\newcommand\RETgcc{118617}
\newcommand\RETmcf{94}
\newcommand\RETmilc{3531}
\newcommand\RETgobmk{22880}
\newcommand\REThmmer{5145}
\newcommand\RETsjeng{1368}
\newcommand\RETlibquantum{1659}
\newcommand\REThref{5874}
\newcommand\RETlbm{75}
\newcommand\RETsphinx{4958}
\newcommand\RETnginx{15004}
\newcommand\RETmusl{10009}

\newcommand\ENTRYperlbench{1596}
\newcommand\ENTRYbzip{66}
\newcommand\ENTRYgcc{4015}
\newcommand\ENTRYmcf{23}
\newcommand\ENTRYmilc{234}
\newcommand\ENTRYgobmk{2388}
\newcommand\ENTRYhmmer{452}
\newcommand\ENTRYsjeng{129}
\newcommand\ENTRYlibquantum{97}
\newcommand\ENTRYhref{508}
\newcommand\ENTRYlbm{16}
\newcommand\ENTRYsphinx{308}
\newcommand\ENTRYnginx{1497}
\newcommand\ENTRYmusl{4400}

\newcommand\PATCHperlbench{62430}
\newcommand\PATCHbzip{896}
\newcommand\PATCHgcc{169067}
\newcommand\PATCHmcf{208}
\newcommand\PATCHmilc{7256}
\newcommand\PATCHgobmk{41632}
\newcommand\PATCHhmmer{9925}
\newcommand\PATCHsjeng{5418}
\newcommand\PATCHlibquantum{1424}
\newcommand\PATCHhref{14824}
\newcommand\PATCHlbm{260}
\newcommand\PATCHsphinx{8010}
\newcommand\PATCHnginx{18984}
\newcommand\PATCHmusl{7594}

\newcommand\TOTALperlbench{\the\numexpr\RETperlbench+\ENTRYperlbench+\PATCHperlbench\relax}
\newcommand\TOTALbzip{\the\numexpr\RETbzip+\ENTRYbzip+\PATCHbzip\relax}
\newcommand\TOTALgcc{\the\numexpr\RETgcc+\ENTRYgcc+\PATCHgcc\relax}
\newcommand\TOTALmcf{\the\numexpr\RETmcf+\ENTRYmcf+\PATCHmcf\relax}
\newcommand\TOTALmilc{\the\numexpr\RETmilc+\ENTRYmilc+\PATCHmilc\relax}
\newcommand\TOTALgobmk{\the\numexpr\RETgobmk+\ENTRYgobmk+\PATCHgobmk\relax}
\newcommand\TOTALhmmer{\the\numexpr\REThmmer+\ENTRYhmmer+\PATCHhmmer\relax}
\newcommand\TOTALsjeng{\the\numexpr\RETsjeng+\ENTRYsjeng+\PATCHsjeng\relax}
\newcommand\TOTALlibquantum{\the\numexpr\RETlibquantum+\ENTRYlibquantum+\PATCHlibquantum\relax}
\newcommand\TOTALhref{\the\numexpr\REThref+\ENTRYhref+\PATCHhref\relax}
\newcommand\TOTALlbm{\the\numexpr\RETlbm+\ENTRYlbm+\PATCHlbm\relax}
\newcommand\TOTALsphinx{\the\numexpr\RETsphinx+\ENTRYsphinx+\PATCHsphinx\relax}
\newcommand\TOTALnginx{\the\numexpr\RETnginx+\ENTRYnginx+\PATCHnginx\relax}
\newcommand\TOTALmusl{\the\numexpr\RETmusl+\ENTRYmusl+\PATCHmusl\relax}

\newcommand\RESperlbench{1115136}
\newcommand\RESbzip{17568}
\newcommand\RESgcc{3074672}
\newcommand\RESmcf{1824}
\newcommand\RESmilc{110688}
\newcommand\RESgobmk{726176}
\newcommand\REShmmer{139216}
\newcommand\RESsjeng{58912}
\newcommand\RESlibquantum{25952}
\newcommand\REShref{278240}
\newcommand\RESlbm{1920}
\newcommand\RESsphinx{103920}
\newcommand\RESnginx{416736}
\newcommand\RESmusl{192153}

\newcommand\METAperlbench{2607559}
\newcommand\METAbzip{78727}
\newcommand\METAgcc{6276870}
\newcommand\METAmcf{19056}
\newcommand\METAmilc{313620}
\newcommand\METAgobmk{3085208}
\newcommand\METAhmmer{574446}
\newcommand\METAsjeng{250234}
\newcommand\METAlibquantum{93222}
\newcommand\METAhref{714629}
\newcommand\METAlbm{16549}
\newcommand\METAsphinx{409814}
\newcommand\METAnginx{1309708}
\newcommand\METAmusl{1238071}

\newcommand\TOTALIIperlbench{\the\numexpr\RESperlbench+\METAperlbench\relax}
\newcommand\TOTALIIbzip{\the\numexpr\RESbzip+\METAbzip\relax}
\newcommand\TOTALIIgcc{\the\numexpr\RESgcc+\METAgcc\relax}
\newcommand\TOTALIImcf{\the\numexpr\RESmcf+\METAmcf\relax}
\newcommand\TOTALIImilc{\the\numexpr\RESmilc+\METAmilc\relax}
\newcommand\TOTALIIgobmk{\the\numexpr\RESgobmk+\METAgobmk\relax}
\newcommand\TOTALIIhmmer{\the\numexpr\REShmmer+\METAhmmer\relax}
\newcommand\TOTALIIsjeng{\the\numexpr\RESsjeng+\METAsjeng\relax}
\newcommand\TOTALIIlibquantum{\the\numexpr\RESlibquantum+\METAlibquantum\relax}
\newcommand\TOTALIIhref{\the\numexpr\REShref+\METAhref\relax}
\newcommand\TOTALIIlbm{\the\numexpr\RESlbm+\METAlbm\relax}
\newcommand\TOTALIIsphinx{\the\numexpr\RESsphinx+\METAsphinx\relax}
\newcommand\TOTALIInginx{\the\numexpr\RESnginx+\METAnginx\relax}
\newcommand\TOTALIImusl{\the\numexpr\RESmusl+\METAmusl\relax}

\definecolor{applegreen}{rgb}{0.55, 0.71, 0.0}
\definecolor{cadmiumgreen}{rgb}{0.0, 0.42, 0.24}

\newcommand{\cmark}{\tikz{\draw[cadmiumgreen,fill=cadmiumgreen] (0,0) circle (1ex);}}%
\newcommand{\xmark}{\small\color{red}$\bm{\times}$}%
\newcommand{\qmark}{\tikz{\filldraw[draw=orange,fill=orange] (0,0) -- (0.2cm,0) -- (0.1cm,0.2cm);}}%
\newcommand{\goodmark}{$\color{blue}\checkmark$}%
\newcommand{\grayxmark}{\small$\bm{\times}$}%
\newcommand{\attack}{\cellcolor{lightgray}}
\newcommand\spacebelowcaption{\vspace{-1.5em}}

\def\BibTeX{{\rm B\kern-.05em{\sc i\kern-.025em b}\kern-.08em
    T\kern-.1667em\lower.7ex\hbox{E}\kern-.125emX}}

\begin{document}
%-------------------------------------------------------------------------------

\date{}
\title{\Large \bf Making Code Re-randomization Practical with \sys}

\author{
{\rm Christopher Jelesnianski}\\
\emph{Virginia Tech}
\and
{\rm Jinwoo Yom}\\
\emph{Virginia Tech}
\and
{\rm Changwoo Min}\\
\emph{Virginia Tech}
\and
{\rm Yeongjin Jang}\\
\emph{Oregon State University}
}

\maketitle

\begin{abstract}
Defense techniques such as Data Execution Prevention (DEP) and Address
Space Layout Randomization (ASLR) were the early role models
preventing primitive code injection and return-oriented programming
(ROP) attacks.
Notably, these techniques did so in an elegant and utilitarian manner,
keeping performance and scalability in the forefront, making them one of
the few widely-adopted defense techniques.
As code re-use has evolved in complexity from JIT-ROP, to BROP and
data-only attacks, defense techniques seem to have tunneled on
defending at all costs, losing-their-way in pragmatic defense design.
Some fail to provide comprehensive coverage, being too narrow in
scope, while others provide unrealistic overheads leaving users
willing to take their chances to maintain performance expectations.

We present \sys, an \emph{on-demand system-wide re-randomization}
technique that improves re-randomization and refocuses efforts to
simultaneously embrace key characteristics of defense techniques:
security, performance, and scalability.
Our code sharing \emph{with} diversification is achieved by implementing
reactive and scalable, rather than continuous or one-time
diversification while the use of hardware supported eXecute-only Memory
(XoM) and shadow stack prevent memory disclosure; entwining and
enabling code sharing further minimizes needed tracking, patching
costs, and memory overhead.
\sys's evaluation shows performance and scalability to have low average
overhead in both compute-intensive (5.5\% on SPEC) and real-world
applications (4.4\% on NGINX).
With this design, \sys demonstrates that strong and scalable security
guarantees are possible to achieve at a practical cost to encourage
deployment.
\end{abstract}

\section{Introduction}
\label{s:intro}

Present day computing continues to trudge through a challenging jungle
of memory corruption vulnerabilities with no clear endgame in sight.
Early in the journey through the attack landscape jungle,
code injection attacks~\cite{aleph:stack-smash,kaempf:heap-vudo,heap-free}
were subsided with the introduction of simplistic yet effective techniques
like Data Execution Prevention (DEP)~\cite{lwn:dep,windows:dep}.

This quickly refocused efforts against stronger adversaries such as code re-use.
Code re-use attacks,
like \emph{return-oriented programming (ROP)}~\cite{shacham:rop} and
ret-into-libc~\cite{shacham:rintolibc},
utilize a victim's program code against itself.
Innocent code snippets, called \emph{gadgets},
are repurposed to construct \emph{gadget-chains},
the equivalent of a malicious payload and initiated via memory corruption vulnerabilities.
To combat code weaponization
and locating these \emph{gadgets} in the first place,
\emph{coarse-grained ASLR}~\cite{pax-aslr,
  addressobf-bhatkar-sec03, addrobfuscat-bhatkar-sec05, kil:aslp},
was created to obstruct the base address of an executable code section
while keeping all code intact.
Because of ASLR's cost-to-benefit ratio and upholding of code sharing,
this pushed for position-independent code to become
the default compilation method for applications to utilize it.

Finding that coarse-grained ASLR could be bypassed by
a single memory disclosure vulnerability,
ASLR moved to \emph{fine-grained randomization} via
randomizing the executable code region
on the scale of instructions~\cite{hiser:ilr, koo:juggling},
basic blocks~\cite{pappas:smashing, davi:isomeron},
or memory pages~\cite{backes:oxymoron}.
However, instruction displacement~\cite{koo:juggling},
as well as in-place code randomization (IPR)~\cite{pappas:smashing},
do not have full coverage of all gadgets present,
leaving some gadgets completely exposed.
Oxymoron~\cite{backes:oxymoron} focused on protecting code pointers
replacing all code references with unique labels and accessing functions via
indexing into a protected table (RaTTle).
To note, Oxymoron is one of the few defenses that is scalable,
supporting code sharing and
not using background processes for tracking or patching code pointers.

Even if code was randomized at very fine granularity,
indirect memory disclosures (via stack and heap) could still be used to
reveal code pages during runtime as in
\emph{just-in-time return-oriented programming (JIT-ROP)},
leaving fine-grained ASLR ineffective.
By collecting addresses and
repeatedly exploiting memory disclosure vulnerabilities, JIT-ROP
spring boards the attacker to reveal
the entire valid executable code region.
Additionally,
Snow \etal~\cite{snow:jitrop} presented a JIT-ROP
capable of reading JIT'ed code within
an application boundary and construct ROP gadget chains on the fly.
Isomeron~\cite{davi:isomeron} was
the only fine-grained ASLR technique capable of preventing
both traditional ROP as well as JIT-ROP,
by randomly switching execution between
two versions of program code
making it unpredictable which version will be chosen.

\emph{eXecute-only Memory (XoM)} was then implemented
to more thoroughly guard memory and restrict access to finding gadgets.
XoM was either ``resilient''
(\eg, Heisenbyte~\cite{tang:heisenbyte}) via destructive code reads
or ``resistant'',
completely preventing reading of the code region
as in Readactor~\cite{crane:readactor}
by using Extended Page Tables (EPT).
While effective against JIT-ROP,
XoM at that time was in its infancy.
Tagging and virtualizing memory via the use of EPT made
any memory access expensive and hog system resources,
leaving fewer resources for the user.
It would not be until recently that
XoM would have hardware support via
Intel Memory Protection Keys (MPK)~\cite{intel:sdm} and
similar protection in ARMv7-M~\cite{arm-xom}.
Though code could not be directly read via XoM,
it could still be deduced via
\emph{inference attacks and de-randomization techniques}.
BROP~\cite{bittau:brop} utilized generalized stack crash reading to
de-randomize ASLR and leak enough gadgets to launch a code reuse attack.
Other code inference attacks involve
crash-less reads~\cite{goktas:information-hiding,gawlik:crop},
allocation oracles~\cite{oikonomopoulos:pokingcpi}, or
zombie gadgets via shared code reloading~\cite{zombie-gadget-snow-oakland16}.
These attacks showed that
randomizing once is simply not sufficient even if protected under XoM;
code remains static thereafter and still vulnerable to any memory leak.

At this point, attacks became more fierce, intimidating defense techniques to
defend by any means necessary and get out of the jungle alive.
\emph{Re-randomization} enables code to become
too volatile to
reliably craft attacks, rendering any leaked knowledge or weaponized code
stale and incorrect, thus thwarting attacks.
Currently, it is assumed that most attacks are carried out remotely or
require I/O system calls to engage.
To counter this, some works opt to trigger via a time interval, ideally
shorter than the network round-trip
latency~\cite{williams:shuffler,chen:codearmor,morpheus-gallagher-asplos19},
while others use known ``code re-use relevant'' events
such as \cc{fork()}~\cite{lu:runtimeASLR} or
I/O system calls~\cite{bigelow:tasr,wang:reranz}.
However,
this assumption does not paint the entire picture of the ROP attack landscape.
Defenses that utilize a \emph{threshold} whether via time,
such as Shuffler~\cite{williams:shuffler} and CodeArmor~\cite{chen:codearmor},
or via leaked-bytes as in ReRanz~\cite{wang:reranz},
are vulnerable to inevitably faster low-profile attacks as
attacks continue to evolve.
Therefore it can be reasoned that
using an interval is simply not a comprehensive metric and
that the concept of using a \emph{threshold} is a liability.

TASR~\cite{bigelow:tasr} does not rely on a threshold,
expecting attackers usage of I/O system calls like \cc{write()}.
While they do prevent remote JIT-ROP that
use I/O system calls,
TASR cannot prevent memory disclosure within
the application boundary~\cite{rudd:aocr}.
These re-randomization techniques forgo expensive
execute-only memory, and instead pair randomization with
protecting code pointers.
Early re-randomization works such as
RuntimeASLR~\cite{lu:runtimeASLR} and TASR~\cite{bigelow:tasr} use
mutable code pointer approaches
to ideally perturb a significant amount of live code,
but this comes with an equally significant performance cost
(\eg, 30-50\% in TASR~\cite{wang:reranz,williams:shuffler})
associated with pointer tracking and patching at
re-randomiztion, severely limiting the effective re-randomization
frequency possible.

Immutable code pointer approaches such as
CodeArmor, Shuffler, and ReRanz are much more lightweight
in terms of tracking and patching.
Although CodeArmor uses segmentation (e.g., \cc{\%gs}) with offsetting,
this technique still allows for
ROP gadgets as well as function pointers to be reached
(\eg, \cc{f+o == \%gs:f+o},
where \cc{f} is an immutable code pointer and \cc{o} is an offset).
Trampoline and indirection tables,
used by Shuffler and ReRanz, respectively,
constrain this loophole to only full-function code reuse
(\eg, allow only the case of \cc{f+o == f'},
where \cc{f'} is another function in the trampoline);
completely eliminating full-function code reuse and
data-oriented programming~\cite{hu-dop} is ongoing research.

Regrettably,
no current re-randomization techniques
take into consideration the scalability of their approaches;
support for code sharing has been forgotten,
and the prevalence of multi-core has excused
the reliance on per-process background threads.
In short,
re-randomization is not as impenetrable and competitive as initially thought.
Current work has shown that making
a secure, practical, and scalable ROP defense technique is challenging.
Even if recent defenses have made some headway through the jungle,
most still lack effective comprehensiveness in security for the system
resource demands they require in return (both CPU and memory);
these factors are prime showstoppers for deployment.

In this paper,
we introduce \sys to refocus defense technique design, showing that it is possible to embrace the core fundamentals of
performance and scalability, while ensuring comprehensive security guarantees.
\sys builds on insight that
thresholds like time intervals as in Shuffler and CodeArmor or
leaked-data-amount in ReRanz are a security loophole and a performance shackle in re-randomization;
\sys does not rely on a threshold whatsoever in its design.
This lets \sys completely side-step
the no-win trade-off between security and performance.
Instead, \sys borrows
the event trigger design but pairs it with XoM violations.
Using XoM, \sys provides complete prevention of JIT-ROP
protecting against \emph{both} variations of remote JIT-ROP as well as
local JIT-ROP,
compared to TASR which can only defend against the former,
with almost zero overhead by using Intel MPK XoM.
\sys also combines XoM with trampolines by
covering them from read access while
also completely decoupling the function entry and the function body in memory;
unlike in CodeArmor,
this makes it impossible to infer and obtain
ROP gadgets in the middle of a function
from a leaked code pointer.

\sys keeps performance and scalability at its forefront.
\sys does not require expensive code pointer tracking and patching like TASR,
nor does \sys incur significant overhead from continuous re-randomization
triggered by overconservative time intervals or benign I/O system calls as in
Shuffler and ReRanz, respectively.
Additionally, while TASR shows a very practical average overhead of 2.1\%,
it has been reported by
Shuffler~\cite{williams:shuffler} and ReRanz~\cite{wang:reranz} that
TASR's overhead against a more realistic baseline
(not using compiler flag \cc{-Og}) is closer to 30-50\% overhead.
Finally, \sys is designed to both support code sharing and not require
the use of any additional system resources (\eg, background threads as
used in numerous works~\cite{williams:shuffler, chen:codearmor,
bigelow:tasr, wang:reranz, morpheus-gallagher-asplos19}) with the help
of Linux kernel memory management and leveraging its own calling
convention.
To summarize, this paper makes the following contributions:
\squishlists
%---------------------------------------------------------
\item \PP{ROP attack \& defense analysis}
Our background~\autoref{s:bg} describes the four prevalent ROP attacks that
challenge current works, including JIT-ROP, code-inference, low-profile,
and code pointer offsetting attacks.
In addition, we describe the bottom-line security implications for each attack.
With this, we classify and exhibit current state-of-the-art defenses standings
on three fronts: security, performance, and scalability.
Our findings show most defenses are not as secure or as practical as expected
against current ROP attack variants.

%---------------------------------------------------------
\item \vspace{2px}
\noindent{\bf \sys defense framework.}
We present the design of \sys in~\autoref{s:design}, a comprehensive ROP defense technique
capable of addressing all currently known ROP attacks.

%---------------------------------------------------------
\item \PP{Scalability and shared code support}
\sys creates and uses a new calling convention in order to be able to
both leverage a shadow stack and minimize the overhead of pointer tracking.
This calling convention is also what enables shared code (\eg, libraries) to
be even more secure, able to be re-randomized by any host process and
maintain the integrity for the rest of the entire system. To the best of our
knowledge, \sys is the first framework capable of this.

%---------------------------------------------------------
\item \PP{Evaluation \& open source prototype}
We implement a prototype of \sys based on LLVM and Linux kernel.
We evaluate and analyze \sys in~\autoref{s:eval} with compute-intensive benchmarks
as well as real-world applications.
\sys's overhead for
compute-intensive benchmarks is 5.5\% on average (geometric
mean) and its worst-case overhead is 18.3\%.
We will open source \sys for the community to
explore other defense mechanisms and build upon our work.
\squishend
\section{Code Layout (Re-)Randomization}
\label{s:bg}

\densetbl
\begin{table*}
  \ra{.9}
  \centering
  \scriptsize
    \begin{tabular}{llccccc|ccc|cc}
  \toprule

    % Something like this..
    % Add local attacker here?
    % ----------------------------------------------------------------------------
    % |  defense  |  Resilience against Attacks |  Meet goals                    |
    % |           |  A1 |  A2 |  A3 |  A4 |  A5 | Per. | Share | Add.res. | long |
    % ============================================================================
    % |coarse-ASLR|  X  |  X  |  X  |  X  |  X  |   O  |   O   |    O     |  X   |
    % |fine-ASLR  |  O  |  X  |  X  |  X  |  X  |   O  |   X   |    O     |  X   |
    % |Oxymoron   |  O  |  X  |  X  |  X  |  X  |   O  |   O   |    O     |  X   |
    % |XOM        |  O  |  O  |  X  |  X  |  X  |   O  |  O/X  |    O     |  X   |
    % |Re-random  |  O  |  O  |  X  |  O  |  X  |   X  |   X   |    X     |  O   |
    % |Mardu      |  O  |  O  |  X  |  O  |  O  |   O  |   O   |    O     |  O   |
    % ----------------------------------------------------------------------------
    % Also adds some specific variants here...

  % first line
  \multirow{2}{*}{\textbf{Types}} &
  \multirow{2}{*}{\textbf{Defenses}} &
  \multicolumn{5}{c}{\textbf{Security}} &
  \multicolumn{3}{c}{\textbf{Performance}} &
  \multicolumn{2}{c}{\textbf{Scalability}}
  \\
  \cmidrule(r){3-12}

  % second line
  &
  &
  \textbf{Gran.} &
  \textbf{A1} &
  % \textbf{A2} &
  \textbf{A2} &
  \textbf{A3} &
  \textbf{A4} &
  \textbf{Perf.} &
  \textbf{Avg.} &
  \textbf{Worst} &
  \textbf{Code Sharing} &
  \textbf{No Addi. Process}
  %\textbf{Longterm}
  \\
  %\hline
  \midrule
  %\hline
  \multirow{3}{*}{\textbf{Load-time ASLR}}&
  Fine-ASLR~\cite{kil:aslp,
  hiser:ilr,pappas:smashing,wartell:stirring,
  conti:selfrando,librando-homescu-ccs13,koo:juggling} &
  Fine &
  \attack\xmark &
  % \xmark &
  \xmark &
  N/A &
  \xmark &
  \goodmark &
  0.4\% &  %actually 0.36%
  6.4\% &  %actually 6.38%
  \grayxmark &
  \goodmark
  %&
  %\xmark
  \\
  &
  Oxymoron~\cite{backes:oxymoron} &
  Coarse &
  \attack\xmark &
  % \xmark &
  \xmark &
  N/A &
  \xmark &
  \goodmark &
  2.7\% &
  11\% &
  \goodmark &
  \goodmark
  %&
  %\xmark
  \\
  &
  Isomeron~\cite{davi:isomeron} &
  Fine &
  \attack\cmark &
  % \xmark &
  \cmark &
  N/A &
  \qmark &      % can reuse the entire function but not ROP
  \grayxmark &
  19\% &
  {\color{red} 42\%} &
  %\XXX{footnote not working} \footnote{This is compared to a PIN instrumented version of SPEC, not vanilla runtime}  &
  \grayxmark &
  \grayxmark
  %&
  %\xmark
  \\
  \midrule
  \multirow{3}{*}{\textbf{Load-time+XoM}} &
  Readactor/Readactor++~\cite{crane:readactor,crane:readactor++} &
  Fine &
  \cmark &
  % \attack\xmark &
  \attack\xmark &
  N/A &
  \qmark &  % trampoline
  \goodmark &
  8.4\% &
  {\color{red} 25\%} &
  \grayxmark &
  \goodmark
  %&
  %\xmark
  \\
  &
  LR 2~\cite{braden:lr2} &
  Fine &
  \cmark &
  % \attack\xmark &
  \attack\xmark &
  N/A &
  \qmark &    % trampoline
  \goodmark &
  6.6\% &
  {18\%} &
  \grayxmark &
  \goodmark
  \\
  &
  kR\^{} X~\cite{pomonis:krx} &
  Fine &
  \cmark &
  % \attack\xmark &
  \attack\xmark &
  N/A &
  \qmark &  % includes phantom function; code ptr -> execute jmp func; -> function
  \goodmark &
  2.32\% &
  {12.1\%} &
  \grayxmark &
  \goodmark
  \\
  \midrule
  \multirow{6}{*}{\textbf{Re-randomization}}
  &
  RuntimeASLR~\cite{lu:runtimeASLR}&
  Coarse &
  \xmark &
  % \xmark &
  \cmark &
  \attack N/A &
  \attack\xmark &
  \grayxmark &
  N/T &
  N/T &
  %\footnote{Did not report any C SPEC benchmarks, only server throughput and latency}
  \goodmark &
  \goodmark
  %&
  %\xmark
  \\
  \cmidrule(r){2-12}
  &
  TASR~\cite{bigelow:tasr} &
  Coarse &
  \qmark &
  % \xmark &
  \cmark &
  \attack\xmark &
  \attack\xmark &
  \grayxmark &
  {\color{red} 2.1\%\textsuperscript{$\dagger$}} &
  {\color{red} 10.1\%\textsuperscript{$\dagger$}} &
  \grayxmark &
  \grayxmark
  %&
  %\xmark
  \\
  &
  ReRanz~\cite{wang:reranz} &
  Fine &
  \qmark &
  % \xmark &
  \cmark &
  \attack\xmark &
  \attack\qmark & % f+o = f'
  \goodmark &
  5.3\% &
  14.4\% &
  \grayxmark &
  \grayxmark
  %&
  %\xmark
  \\
  \cmidrule(r){2-12}
  &
  Shuffler~\cite{williams:shuffler} &
  Fine &
  \cmark &
  % \xmark &
  \cmark &
  \attack\xmark &
  \attack\qmark & % f+o = f'
  \grayxmark &
  14.9\% &
  {\color{red} 40\%} &
  %\footnote{This is worst case for C SPEC, actual worst case is ~45\% from 483.xalancbmk, a C++ SPEC benchmark}
  \grayxmark &
  \grayxmark
  %&
  %\xmark
  \\
  &
  CodeArmor~\cite{chen:codearmor} &
  Coarse &
  \cmark &
  % \xmark &
  \cmark &
  \attack\xmark &
  \attack\xmark &
  \grayxmark &
  3.2\% &
  {\color{red} 55\%} &
  \grayxmark &
  \grayxmark
  %&
  %\xmark
%   \\
%   &
%   Morpheus\XXX{footnote not working}\footnote{Morpheus is a defense implemented on the RISC-V architecture. It is not currently
%   portable to current systems without major hardware support}~\cite{morpheus-gallagher-asplos19}&
%   Fine &
%   \cmark &
%   \qmark &
%   \cmark &
%   \qmark &
%   \cmark &
%   \goodmark &
%   0.84\% &
%   6.71\% &
%   \grayxmark &
%   \grayxmark
%   %&
%   %\goodmark
  \\
  \midrule
  \multirow{1}{*}{\textbf{Our Approach} } &
  \sys &
  Fine &
  \cmark &
  % \qmark &
  \cmark &
  \cmark &
  \qmark & % f+o = f'
  \goodmark &
  5.5\% &
  18.3\% &
  \goodmark &
  \goodmark
  %&
  %\goodmark
  \\

  \bottomrule
\end{tabular}

  \if 0
  Coarse-ASLR~\cite{pax-aslr} &
  \xmark &
  \xmark &
  \xmark &
  \xmark &
  N/A &
  \xmark &
  \xmark &
  \goodmark &
  1\% &
  3\% &
  \cmark &
  \cmark
  %&
  %\xmark
  \\
  &
  \fi

  \caption{
    Classifications of ASLR-based code-reuse defenses.
    Gray highlighting emphasizes the attack ({\it A1-A4}) that largely invalidated each type of defense.
    {\protect\cmark}
    indicates the attack is blocked by the defense. (attack-resistant).
    {\protect\xmark}
    indicates the defense is vulnerable to that attack.
    {\protect\qmark}
    indicates the attack is not blocked but is still mitigated by the
    defense (attack-resilient).
    {\protect\goodmark}
    indicates the defense meets performance/scalability requirements.
    {\protect\grayxmark}
    indicates the defense is unable to meet performance/scalability requirements.
    {\bf N/A} in the column {\it A3} indicates that
    the attack is not applicable to the defense
    due to lacking of re-randomization,
    and {\bf N/T} in the column {\it Performance} indicates that
    either SPEC CPU2006 or \cc{perlbench} is not tested.
    Specifically in the column {\it A1},
    {\protect\qmark} indicates that the defense
    cannot prevent the JIT-ROP attack within the application boundary
    that does not use system calls;
    in the column {\it A4}, {\protect\xmark} indicates that an attack
    may reuse both ROP gadgets and entire functions while {\protect\qmark}
    indicates that an attack can only reuse entire functions.
    {\color{red}\textsuperscript{$\dagger$}} Note that in TASR the
    baseline performance is a binary compiled with \cc{-Og},
    necessary to correctly track code pointers. Previous
    work~\cite{williams:shuffler, wang:reranz} reported
    performance overhead of TASR using regular optimization
    (\cc{-O2}) binary is $\approx$30-50\%.
    \sys provides strong security guarantees with low performance
    overhead and good system-wide scalability compared to existing
    re-randomization approaches.
  }
  \spacebelowcaption
  \vspace{-1em}
  \label{tbl:comparison}
\end{table*}
\densetblend

In this section,
we present a background on
existing code layout re-randomization techniques.
To help understand the attack and defense arms race,
we classify code randomization techniques into
two categories:
load-time randomization and
continuous re-randomization.

To start with,
we describe two attacks,
A1 (JIT-ROP) and A2 (BROP),
aimed to defeat load-time code randomization.
Next,
we describe how continuous re-randomization techniques
defy A1 and A2 by analyzing
design elements of the techniques.
Specifically,
we categorize continuous re-randomization techniques by
its re-randomization triggering condition
(\ie, either based on the system call history or timing threshold)
and the semantics of storing code pointers
(\ie, tracking pointers or use an indirect, immutable function index).
The reason for focusing on those two categories is that
these design elements greatly affect
the security, performance, and scalability of techniques.
Finally,
we compare and contrast each technique
regarding three aspects,
security,
performance,
and scalability.
Particularly for illustrating attack resilience,
we present two attacks,
A3 (low-profile JIT-ROP) and A4 (code pointer offsetting),
to which existing re-randomization techniques are susceptible.
Regarding scalability,
we report the requirement of additional CPU usage for re-randomization and
whether shared code layout is possible among multiple processes.

In summary,
\autoref{tbl:comparison} illustrates the characteristics of
each defense technique by randomization category,
attack resilience, and performance and scalability factors,
and we describe these in detail in the following.

\subsection{Attacks against Load-time Randomization}
\label{s:bg:loadtime}
Load-time code randomization techniques suffer from attacks A1
(JIT-ROP) and A2 (BROP, etc.).
In the following, we describe characteristics of techniques and
attacks.

\subsubsection{Load-time Randomization without XoM}
\label{s:bg:loadtime:woxom}
Code layout randomization techniques so-called coarse-grained
ASLR~\cite{pax-aslr} or fine-grained ASLR~\cite{kil:aslp,
hiser:ilr,pappas:smashing,wartell:stirring,backes:oxymoron,
davi:isomeron,conti:selfrando,librando-homescu-ccs13,koo:juggling},
depending on the granularity of layout randomization, fall into this
category of code layout randomization.
These techniques randomize the code layout only once, usually when
code is loaded into memory.
After code is loaded and shuffled, its layout never changes
during the lifetime of the program.
The following attack can defeat the security guarantee of load-time
randomization techniques.

\PP{A1: Just-in-time ROP (JIT-ROP)}
An attacker with arbitrary memory read capability
may launch JIT-ROP~\cite{snow:jitrop} by interactively performing
memory reads to disclose one code pointer.
This disclosure can be used to then leap frog and
further disclose other addresses
to ultimately learn the code contents in memory.
Any load-time code randomization technique
that does not protect code from read access
including fine-grained ASLR techniques
is susceptible to this attack.

\boxbeg
{\bf Security implications:}
Techniques failing to protect code from
read
access
allows
code-reuse attacks to be launched
regardless of code
randomization granularity.
\boxend

\subsubsection{Load-time Randomization with XoM}
\label{s:bg:loadtime:wxom}
In response to A1 (JIT-ROP),
several research projects protect code from read access
via
destructive read memory~\cite{tang:heisenbyte} or
execute-only memory~\cite{backes:xnr,
gionta:hidem,
crane:readactor,
crane:readactor++,
tang:heisenbyte,
braden:lr2,
werner:near,
pomonis:krx,
chen:codearmor}.
Systems with destructive read~\cite{tang:heisenbyte}
allow code pointer leaks to occur, but trigger
intended localized code corruption once a read attempt is made, such that
code-reuse attacks following the read
will fail to execute the expected code by the attacker.
Systems with execute-only memory (XoM)~\cite{backes:xnr,
gionta:hidem,
crane:readactor,
crane:readactor++,
tang:heisenbyte,
braden:lr2,
werner:near,
pomonis:krx,
chen:codearmor}
aim to fundamentally block all read attempts
of program code by
removing read permissions from the code area.
Applying these techniques prevent attackers from
gaining knowledge about code contents,
and thereby, nullifying A1.
However, leaving the code layout fixed
after load-time randomization makes
these techniques susceptible to
the following attack.

\PP{A2: Blind ROP (BROP) and code inference attacks}
Even with protecting code from read access (\ie, XoM),
load-time randomizations still
are susceptible to BROP~\cite{bittau:brop}
and/or other code content inference
attacks~\cite{zombie-gadget-snow-oakland16, bgdx-pewny-acsac17}.
Although these attacks do not read code directly,
attackers may accumulate information about the code contents
by conducting probing on the code many times
because code layout will never change after it is loaded and shuffled.
In particular,
BROP is a clone-probing attack that infers code contents via
observing differences in execution behaviors such as
timing or program output.
Other attacks~\cite{zombie-gadget-snow-oakland16, bgdx-pewny-acsac17}
defeat destructive code read defenses~\cite{tang:heisenbyte,
werner:near}
by inferring code contents from
a small fraction of a code read and then
weaponizing inferred code.

\boxbeg
{\bf Security implications:}
Maintaining a fixed layout over crash-probing or read access to code
allows inferring code contents indirectly, and thereby,
attackers can still learn the code layout and launch code-reuse attacks.
\boxend

\subsection{Continuous Re-randomization Defeating A1 \& A2}
\label{s:bg:rerand-intro}

In response to A1 (JIT-ROP) and A2 (BROP, etc.),
continuous re-randomization techniques~\cite{giuffrida:asr3, lu:runtimeASLR,
bigelow:tasr, williams:shuffler, chen:codearmor, wang:reranz,
morpheus-gallagher-asplos19}
aim to defeat attacks by
continuously shuffling code (and data) layouts at runtime
to make information (code or code addresses) leaks or
code probing done before shuffling useless.

To illustrate the internals of
re-randomization techniques in a nutshell,
we describe the core design elements of re-randomization
by categorizing them into two,
based on their design elements:
{\it 1) Re-randomization triggering condition} and
{\it 2) Code pointer semantics}.

\PP{Re-randomization triggering condition}
Existing continuous re-randomization techniques
trigger their randomization
based on the following two conditions.

\squishlists
\item {\bf Timing:}
    Techniques~\cite{williams:shuffler, chen:codearmor}
    shuffle the layout periodically by
    setting a timing window for layout randomization.
    For example, Shuffler~\cite{williams:shuffler} triggers
    re-randomization every 50~msec,
    and CodeArmor~\cite{chen:codearmor} can set
    re-randomization period
    as low as 55~$\mu$sec.

\item {\bf System-call history:}
  Techniques~\cite{lu:runtimeASLR, bigelow:tasr, wang:reranz}
  shuffle the layout based on
  the history of the program's previous system call invocations,
  \eg, after invoking \cc{fork()}~\cite{lu:runtimeASLR} or
  when \cc{write()} (leak) is followed by
  \cc{read()} (exploit)~\cite{bigelow:tasr, wang:reranz}.
\squishend

\PP{Code pointer semantics}
Existing continuous re-randomization techniques
use the following three different types of
code pointer semantics.

\squishlists
\item {\bf Code address as code pointer:}
Code pointers store
the actual addresses.
In this case,
leaking a code pointer lets
the attacker have knowledge about
code address.
Therefore,
techniques in this category~\cite{giuffrida:asr3, bigelow:tasr} require
tracking of code pointers (or all pointers) at runtime,
which is computation expensive,
to update their values after
randomizing the code (and data) layout.
For instance, TASR~\cite{bigelow:tasr} shows very high performance
overhead (30-50\%) especially for I/O-intensive applications, such as
web servers~\cite{williams:shuffler, wang:reranz}.

\item {\bf Function trampoline address as code pointer:}
Code pointers store
a function table index~\cite{williams:shuffler} or
the address of a function trampoline~\cite{wang:reranz}.
This design avoids the expensive pointer tracking
in order to enhance the performance of
re-randomization techniques.
Instead of tracking and updating code pointers,
techniques in this category
setup a function table,
which stores all function addresses of the program,
and store an index of the table in the code pointer
to refer a function.
After re-randomization and re-locating the code layout,
the techniques update only the function table
while all code pointers remain immutable.
With this design,
leaking a code pointer will tell the attacker
only the semantics of referring to a function
(\ie, function index in the trampoline)
but not about the code layout.

\item {\bf An offset to the code address as code pointer:}
Code pointers store an offset from
the (randomized) base address of the layout.
This design is also intended for avoiding pointer tracking
by having an immutable offset from the random version address for
referring to a function,
as in CodeArmor~\cite{chen:codearmor}.
At re-randomization,
updating code layout only requires
updating the random version base address,
and does not require any update of pointers.
With this design,
leaking a code pointer will tell the attacker
the offset to select a function
no matter how the code layout is randomized.
\squishend

\subsection{Attacks against Continuous Re-randomization}
\label{s:bg:rerand-attack}

Continuous re-randomization techniques suffer from
two attacks (A3 and A4) that we define in this section.

\noindent{\bf A3: Low-profile JIT-ROP.}
This is class of attacks 
does not trigger re-randomization either
by completing the attack quickly or
without requiring I/O system calls.
As the trigger for layout re-randomization,
Existing defenses utilize one of timing~\cite{giuffrida:asr3,
williams:shuffler, chen:codearmor, morpheus-gallagher-asplos19},
amount of transmitted data by output system calls~\cite{wang:reranz},
or I/O system call boundary~\cite{bigelow:tasr}
as the trigger for layout re-randomization.
Therefore attacks within the application boundary,
such as code-reuse attacks in Javascript engine
where both information-leak followed by
control-flow hijacking attack
may conclude faster than the re-randomization timing threshold or
not interact with I/O system call,
can bypass these triggering conditions.
The code layout may remain unchanged within the given interval,
and thereby,
an attacker may launch JIT-ROP to unchanged code layout.

\boxbeg
{\bf Security implications:}
An attack may be completed within
a defenses pre-defined randomization time interval or
without involving any I/O system call invocation.
In such a case,
basing the re-randomization trigger on
timing or system call history
allows attackers
bypass re-randomization and
launch A1 (JIT-ROP) successfully.
\boxend

\PP{A4: Code pointer offsetting}
Even with re-randomization,
techniques might be susceptible to a code pointer offsetting attack
if code pointers are not protected from
having arithmetic operations applied by attackers~\cite{bigelow:tasr,chen:codearmor}.
An attacker may trigger a vulnerability to apply
arithmetic operations to an existing code pointer.
For example,
suppose a code pointer \cc{p} points to a function \cc{f()},
and altering the value of \cc{p} by adding an offset \cc{o}
could make \cc{p} point another code address \cc{p+o}.
Particularly, in techniques directly using code
address~\cite{bigelow:tasr} or code offset~\cite{chen:codearmor},
\cc{p+o} could be even a ROP gadget in \cc{f()} if the attacker knows
the gadgets offset \cc{o} beforehand.
Ward~\etal~\cite{relrop-wardleakage-esorics19}
has recently demonstrated that this attack is possible against TASR.

\boxbeg
{\bf Security implications:}
Maintaining a fixed code layout across
re-randomizations and
not protecting code pointers
lets attackers perform arithmetic operations over pointers,
allowing launching other
ROP gadgets.
\boxend

\section{Threat Model and Assumptions}
\label{s:threat}
We build \sys based on the following assumptions.

\noindent\textbf{Attacker's capability:}
\squishlists
\item \PP{Arbitrary read/write}
Attackers can perform arbitrary memory read/write (if the address is
readable/writeable) in a target process
by exploiting software vulnerabilities in the victim program.
With this capability, attackers may launch attacks A1--A4
on an unprotected system.

\item \PP{Local brute-force attacks}
We assume that all attack attempts are run in a local machine.
In this regard, attacks may be performed any number of times within
a short time period (\eg, within a millisecond).
This assumption gives the capability of launching attacks A1--A4 without
triggering re-randomization in prior systems~\cite{giuffrida:asr3,
  bigelow:tasr, williams:shuffler, wang:reranz, chen:codearmor}.
\squishend

\noindent\textbf{System and trusted computing base:}
\squishlists
\item \PP{XoM (R$\oplus$X) and DEP (W$\oplus$X)}
We assume that the userspace of the system does not have any memory region that is
both readable and executable.
Likewise, we assume that the userspace of the system does not have any memory
region that is both writable and executable.

\item \PP{Trusted hardware and no physical access}
We assume all hardware is trusted and attackers do not have physical access.
Particularly,
we trust Intel Memory Protection Keys (MPK)~\cite{intel:sdm}, a
mechanism that provides eXecute-Only Memory (XoM),
and we regard attacks to CPU (side-channel attacks, \eg,
Spectre~\cite{spectre-kocher-sp19},
Meltdown~\cite{meltdown-lipp-sec18}) to be out of scope.

\item \PP{Trusted kernel and program loading}
We trust the OS kernel and the loading/linking process of the program
(\cc{execve()}, \cc{ld-linux}, etc.), thus attackers cannot intervene to
perform any attack before the program starts.
\squishend
\section{\sys Design}
\label{s:design}
We begin by the design overview of \sys (\autoref{s:design:ov}) and
then detail \sys compiler (\autoref{s:design-compiler}) and kernel
(\autoref{s:design-kernel}).

\subsection{Overview}
\label{s:design:ov}

This section presents the overview of \sys, along with its design
goals, challenges, and outlines its architecture.

%%%%%%%%%%%%%%%%%%%%%%%%%%
\subsubsection{Goals}
\label{s:design:ov:goal}
Understanding how the attack landscape and existing mitigations fit
together, our goal in designing \sys is to shore up the current
state-of-the-art to enable a practical code randomization.
More specifically, our design goals are as follows:

\PP{Security}
No prior solutions provide a comprehensive defense against existing
attacks (see~\autoref{s:bg}).
Systems with only load-time ASLR are susceptible to leaking
code-content (A1) and letting attackers infer code-content (A2).
Systems applying re-randomization are still susceptible to
low-profile attacks (A3)
and code pointer offsetting attacks (A4).
\sys aims at either defeating or significantly limiting the capability
of attackers in launching code-reuse attacks spanning from A1 to A4 to
provide best-effort security against existing attacks.

\PP{Performance}
Many prior approaches~\cite{crane:readactor,
  crane:readactor++,williams:shuffler,chen:codearmor} demonstrate
decent runtime performance in average cases ($<$ 10~\%, \eg, $<$
3.2~\% in CodeArmor); however, they also show evidence of
a few scenarios that are remarkably slow (\ie, $>$ 55~\%,
see listed numbers in \emph{Worst} column in
\autoref{tbl:comparison}).
We design \sys to run with
an acceptable average overhead ($\approx$ 5~\%) with
minimal performance outliers across a variety of application types.

\PP{Scalability}
Most proposed exploit mitigation mechanisms have overlooked
the impact of required additional system resources,
such as memory or CPU usage,
which we consider a scalability factor.
This is crucial for applying a defense system-wide,
and is even more critical when
deploying the defense in the pay-as-you-go pricing Cloud.
Oxymoron~\cite{backes:oxymoron} is the
only defense that allows code sharing of randomized code. No
advanced re-randomization defenses support code sharing thus they
require significantly more memory.
Additionally, most re-randomization
defenses~\cite{williams:shuffler, wang:reranz, chen:codearmor}
require per-process background threads, which
not only cause additional CPU usage
but also contention with the application process.
As a result, approaches requiring per-process/thread
background threads show significant performance overhead as
the number of processes increases.
For example, Shuffler~\cite{williams:shuffler} shows around 55\%
performance overhead when four NGINX workers (plus four Shuffler
threads) run on two cores with 50 ms shuffling interval.

Therefore, to apply \sys system-wide, we design \sys to not require
significant additional system resources, for instance, additional
processes/threads or significant additional memory.

%%%%%%%%%%%%%%%%%%%%%%%%%%
\subsubsection{Challenges}
\label{s:design:ov:challenge}

It is challenging to achieve all of aforementioned goals. One naive
approach is to mix all good defenses from existing approaches but such
an approach fails to meet the goal because requirements for enabling
each defense could conflict. Hence, we list challenges in achieving
our goals in the following.

\PP{Tradeoffs in security, performance, and scalability}
An example of the tradeoff between security and performance is having
fine-grain ASLR with re-randomization. Although such an approach can
defeat A4, systems cannot apply such protection because the
re-randomization must finish quickly to meet the performance goal and
also defeat A3.
An example of the tradeoff between scalability and performance is
having a dedicated process/thread for performing re-randomization.
However, this results in a drawback in scalability by requiring more
CPU time in the entire system.
Therefore, a good design must find a breakthrough to meet \emph{all}
of aforementioned goals.

\PP{Conflict in code-diversification vs. code-sharing}
Layout re-randomization requires
diversification of code layout per process,
and this affects the availability of code-sharing.
The status quo is that
code sharing cannot be applied to
any existing re-randomization approaches,
and this makes the defense unable to scale to protect
many-process applications.
Although Oxymoron~\cite{backes:oxymoron} enables
both diversification and sharing of code,
it does not consider re-randomization, nor
use a sufficient randomization granularity (page-level),
which is insufficient against A4.

%%%%%%%%%%%%%%%%%%%%%%%%%%%%
\subsubsection{Architecture}
\label{s:design:ov:arch}

\begin{figure}[t!]
  \centering
  \includegraphics[width=\columnwidth]{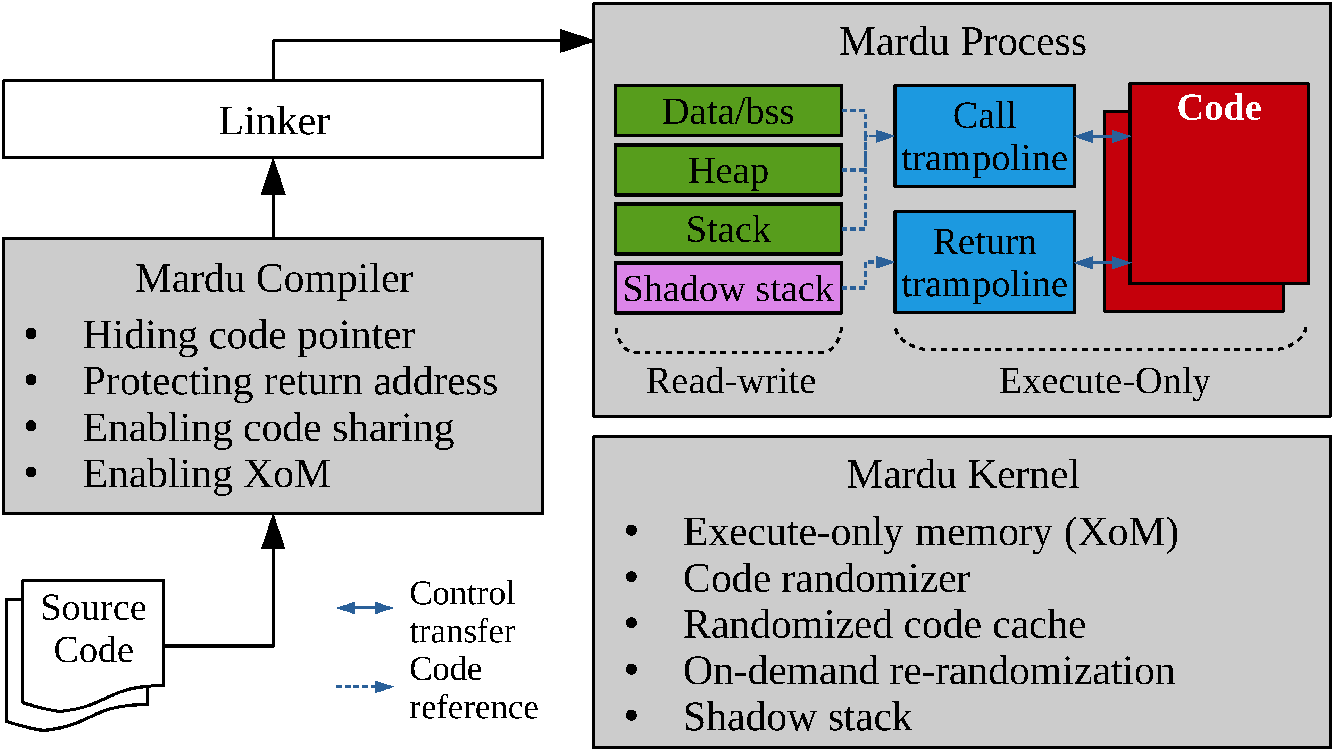}
  \caption{Overview of \sys}
  \spacebelowcaption
 \label{f:overview}
\end{figure}

%%%%%%%%%%%%%%%%%%%%%%%%%%%%%%%%%%%%%%%%%%%%%%%%%%%%%%%%%%%%%%%%%%%%%%
We design \sys to make a breakthrough beyond tradeoffs in security,
performance, and scalability, which satisfies all three aspects and
becomes practical.
We introduce our approach for each aspect below:

\PP{Scalability: Sharing randomized code}
\sys manages the cache of randomized code in the kernel,
which is capable of being
mapped to multiple userspace processes and is not readable
from userspace. Thus, it does not require any additional memory.

\PP{Scalability: System-wide re-randomization}
Since code is shared between processes in \sys, per process randomization,
which is CPU intensive, is not required;
rather a single process randomization is sufficient for the \emph{entire} system.
For example,
once a worker process of NGNIX web server crashes, it re-randomizes
all its mapped executables (\eg, \libc) upon exit.
This re-randomizes all
processes using the same executables (\eg, \libc of all processes,
including other workers of the NGNIX server, will be immediately
re-randomized).

\PP{Scalability: On-demand re-randomization}
\sys re-randomizes code only when suspicious activity is
detected. \sys considers any probing on code memory and program crash
as suspicious activity and efficiently detects it using
eXecute-Only-Memory (XoM). This design is advantageous because \sys
needs neither per-process background threads nor re-randomization
interval unlike prior re-randomization approaches. Particularly, \sys
re-randomization is performed in the context of a crashing process,
thereby not affecting the performance of other normal running
processes.

\PP{Performance: Immutable code pointers}
The above described design decisions for scalability also help
reduce performance overhead.
In addition, \sys neither tracks nor
encrypts code pointers so it does not mutate code pointers upon
re-randomization.
While this design choice minimizes performance overhead, other
security features (\eg, XoM, trampoline, and shadow stack) in \sys
ensure the comprehensive ROP defense.

\PP{Security: Detecting suspicious activities}
\sys considers any process crash or code probing attempt as a
suspicious activity. \sys's use of XoM makes any code probing attempt trigger
process crash and system-wide re-randomization. Therefore, \sys
counters direct memory disclosure attacks as well as code inference
attacks requiring initial code
probing~\cite{zombie-gadget-snow-oakland16, bgdx-pewny-acsac17}. To
implement XoM, we use Intel MPK~\cite{intel:sdm} so our XoM design
does not impose any runtime overhead unlike virtualization-based
designs.

\PP{Security: Preventing code \& code pointer leakage}
In addition to system-wide re-randomization, \sys is designed to
minimize the leakage of code and code pointers. Besides XoM, we use
three techniques.
First, \sys applications always go through a trampoline region to
enter into or return from a function. Thus, only trampoline addresses
are stored in memory (\eg, stack and heap) while non-trampoline code
pointers remain hidden. \sys does not randomize where the trampoline
region is so \sys does not need to track and patch code pointers in
memory upon re-randomization.
Second, \sys performs fine-grained function-level randomization within
an executable (\eg, \libc) to completely disconnect any correlation
between trampoline addresses and code addresses. This provides high
entropy (\ie, roughly $n!$ where $n$ is the number of functions), so
it is not feasible to succeed BROP~\cite{bittau:brop} without any
crash. Also, unlike re-randomization approaches that rely on shifting
code base addresses~\cite{lu:runtimeASLR, bigelow:tasr,
  chen:codearmor}, \sys is not
susceptible to code pointer offsetting attack (A4).
Finally, \sys stores return addresses--precisely, trampoline addresses
for return--in a shadow stack; the shadow stack stores only return
addresses and is hidden under a segmentation register in x86. This
design makes stack pivoting practically infeasible.

\PP{Design overview}
As shown in \autoref{f:overview}, \sys is composed of compiler and
kernel components.
The \sys compiler replaces \call and \ret instructions with functionally
equivalent \jmp's to the trampoline region and generates code to store
return addresses in a shadow stack.
\sys compiler generates
PC-relative code so randomized code can be shared by multiple
processes. Also, the compiler generates and attaches additional metadata to binaries
for efficient patching of PC-relative addressing code upon
(re-)randomization.
The compiler separates data from code pages to
prevent false-positive XoM violations, from \sys applications
attempting to read inter-mixed data in protected code regions.

The \sys kernel is responsible for choreographing the runtime when a
\sys executable is launched. The kernel extracts and loads the
executable's compiler-generated metadata into a cache to be shared by
multiple processes. This data is then used by \sys for first load-time
randomization as well as re-randomization. The randomized code is
cached and shared by multiple processes; while allowing sharing, each
process will get a different random virtual address space for the
shared code.
\sys kernel prevents read operations of the code region including the
trampoline region using XoM so trampoline addresses do not leak
information about non-trampoline code. Whenever a process crashes
(\eg, XoM violation), \sys kernel re-randomizes all associated shared
code so all relevant processes are re-randomized to thwart an
attacker's knowledge immediately.
\subsection{\sys Compiler}
\label{s:design-compiler}

\sys compiler generates a binary able to
1) hide its code pointers,
2) share its randomized code among processes,
and 3) run under XoM.
To this end,
\sys uses its own calling convention using
a trampoline region and shadow stack.

\subsubsection{Code Pointer Hiding}
\label{s:dc:tr-ss}

\begin{figure}[t!]
  \centering
  \includegraphics[width=\columnwidth]{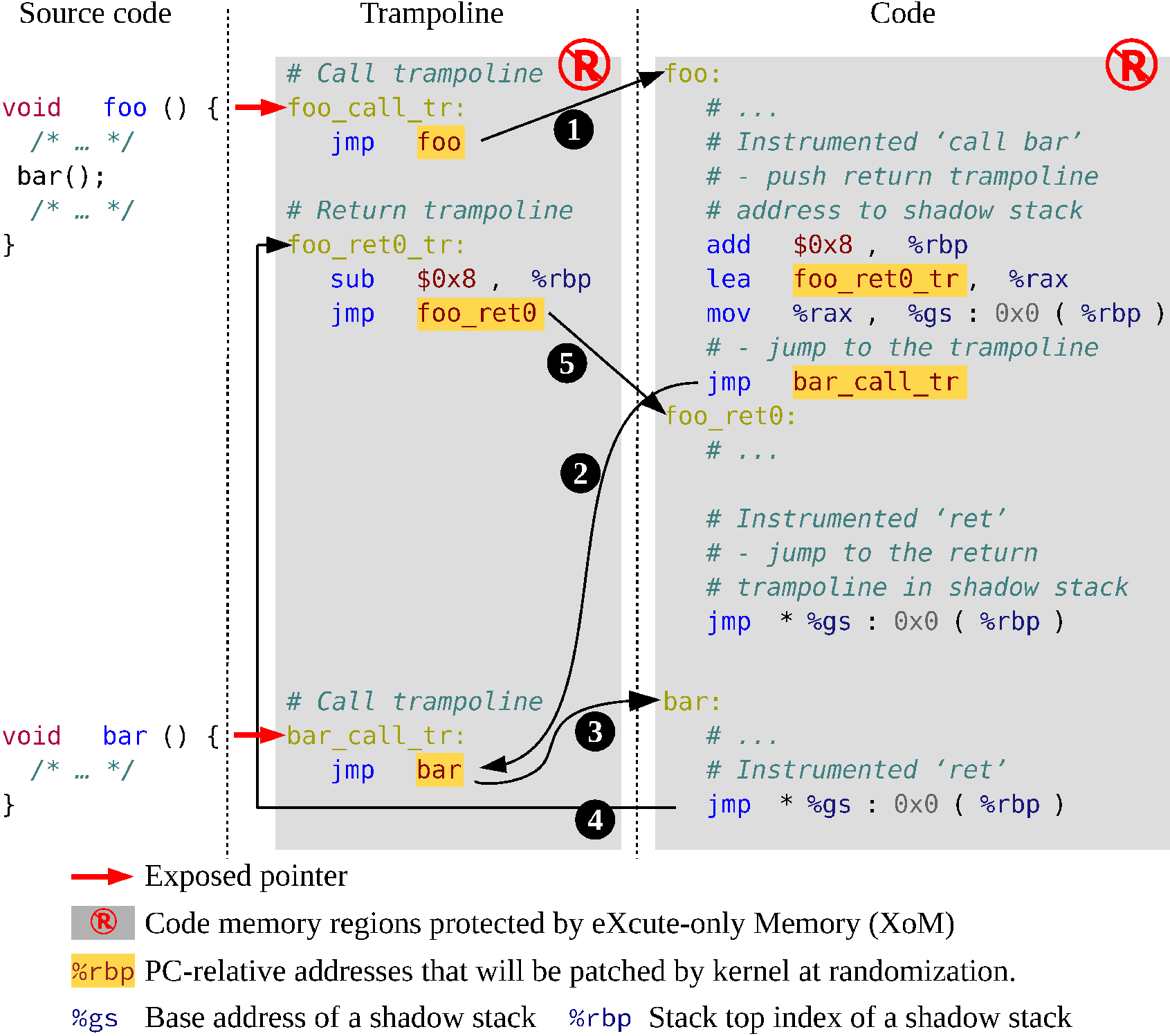}
  \caption{Illustrative example executing a \sys-compiled function
    \fnfoo, which calls a function \fnbar and then returns.
  }
  \spacebelowcaption
\label{f:trampoline}
\end{figure}

\PP{Trampoline}
\sys hides code pointers without
paying for costly runtime code pointer tracking.
The key idea for enabling this is
to split a binary into two regions in process memory:
\emph{trampoline} and \emph{code} regions
(as shown in \autoref{f:trampoline} and \autoref{f:mem}).
A trampoline is an intermediary call site that
moves control flow securely to/from a function body,
protecting the XoM hidden code region.
There are two kinds of trampolines: call and return trampolines.
As their names imply,
a \emph{call trampoline} is responsible for
forwarding control flow from an instrumented call to
the \emph{code region} function entry,
while a \emph{return trampoline} is responsible for
returning control flow semantically to the caller.
Each function has one call trampoline to its function entry,
and each call site has one return trampoline returning to
the following instruction of the caller.
Since trampolines are stationary, \sys
does not need to track
code pointers upon re-randomization because only stationary call trampoline
addresses are exposed to memory.

\PP{Shadow stack}
Unlike the
x86 calling convention using \call/\ret
to store
return addresses on the stack, \sys instead stores all return addresses in a shadow
stack and leaves data destined for the regular stack
untouched.
Effectively, this protects all backward-edges.
An instrumented \call pushes
a return trampoline address to the shadow stack and then
jumps to a call trampoline;
an instrumented \ret directly jumps to the return trampoline address
at the current top of the shadow stack.
The base address of the \sys shadow stack is randomized by ASLR and is hidden in
segment register \gs, which cannot be modified in userspace and will never be
stored in memory.
Therefore, it is infeasible to know the shadow stack base address without
causing a program crash.
We additionally reserve one register, \rbp, to use exclusively
as a stack top index of a shadow stack in order to avoid costly
memory access.

\PP{Running example}
\autoref{f:trampoline} is an example of executing a \sys-compiled
function \fnfoo, which calls a function \fnbar and then returns.
Every function call and return goes through trampoline code which
stores the return address to a shadow stack, of which base address is
hidden in register \gs.
The body of \fnfoo is entered via its call trampoline \protect
\BC{1}. Before \fnfoo calls \fnbar, the return trampoline address is
stored to the shadow stack--each call site has one return trampoline
returning to the next instruction of the call site. Control flow then
jumps to \fnbar's trampoline \protect \BC{2}, which will jump to the
function body of \fnbar \protect \BC{3}. \fnbar returns to the address
in the top of the shadow stack, which is the return trampoline address
\protect \BC{4}. Finally, the return trampoline returns to the
instruction following the call in \fnfoo \protect \BC{5}.

\subsubsection{Enabling Code Sharing among Processes}
\label{s:dc:sharing}

\PP{PC-relative addressing}
To enable sharing, \sys compiler generates PC-relative (\ie,
position-independent) code
so code can be shared amongst processes that load the same code in different
virtual addresses.
The key challenge here is \emph{how to incorporate PC-relative addressing with
randomization}. \sys randomly places code (at function granularity) while
trampoline regions are stationary.  This means any code using PC-relative
addressing must be correspondingly fixed up once its randomized location is
decided.  In \autoref{f:trampoline}, all jump targets between the trampoline
and code, denoted in yellow rectangles, are PC-relative and must be fixed.
Also, all data addressing instructions are PC-relative (\eg, accessing global
data, GOT, \etc) and also must be fixed.

\PP{Fixup information for patching}
With this policy, it is necessary to keep track of
these instructions to patch them properly during runtime. To make the
runtime patching process simple and efficient, \sys compiler generates
additional metadata into the binary that describes exact locations
for patching and their file-relative offset. This fixup information
makes the patching as simple as just adjusting PC-relative offsets for
given locations within analyzing instructions (see
\autoref{f:mem}). Since displacement in PC-relative addressing is
32 bits in size in x86-64 architecture, $\pm2$~GB is the maximum offset
from the \rip supported by this addressing mode.
We elaborate on the patching process in~\autoref{s:dk:rand}.

\PP{Supporting a shared library}
A call to a shared library is treated the same as
a normal function call to preserve the code pointer hiding property;
that is, \sys refers to the call trampoline
for the shared library call via PLT/GOT.
It first calls the PLT (Procedure Linkage Table) via a trampoline,
which jumps to an external function
whose address is not known at link time and
left to be resolved by the dynamic linker.
The result of dynamic symbol resolution
is a function address in the call trampoline of the external library,
and it is stored in a GOT (Global Offset Table) for caching.
While \sys does not specifically protect GOT, we assume that GOT is
already protected using MPK~\cite{mpk-glibc-larabel,
  hardening-fedora}.

\PP{Overhead of sharing}
\sys's sharing mechanism does not have noticeable runtime overhead as
PC-relative code is already mandatory to enable ASLR. In addition, the overhead
of runtime patching is negligible because \sys avoids
``stopping the world'' when patching the code to maintain internal
consistency compared to other approaches.

\subsubsection{Enabling Execute-Only-Memory}
\label{s:dc:xom}
Finally, to run code with XoM, \sys compiler ensures code and data are
segregated in different pages. Compilers sometimes intermingle data
within \cc{.text} code section as an optimization. However, if this data is
attempted to be read during runtime, an XoM violation will be raised.
As previous work~\cite{crane:readactor} reported, we found that Clang
intermingles code and data only for jump tables so we disable
generating jump tables in \cc{.text} section.
\subsection{\sys Kernel}
\label{s:design-kernel}

\begin{figure}[t!]
  \centering
  \includegraphics[width=\columnwidth]{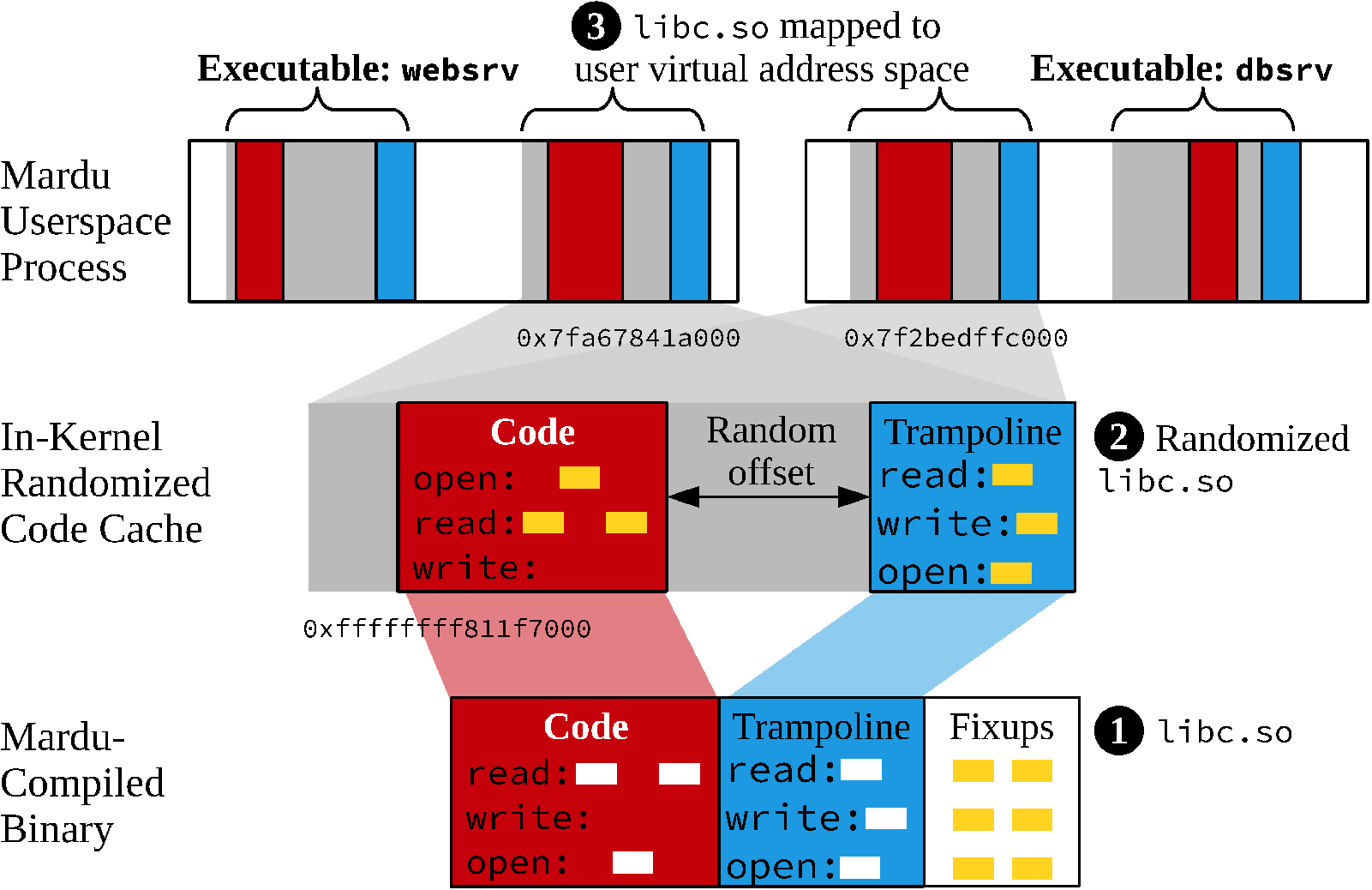}
  \caption{Memory layout of two \sys processes: \cc{websrv} (top left)
    and \cc{dbsrv} (top right).
    The randomized code in kernel (\cc{0xffffffff811f7000}) is
    shared by multiple processes, which is mapped to its own virtual
    base address (\cc{0x7fa67841a000} for \cc{websrv} and
    \cc{0x7f2bedffc000} for \cc{dbsrv}).
  }
  \spacebelowcaption
 \label{f:mem}
\end{figure}

\sys kernel randomizes code at load-time and runtime. It maps
already-randomized code, if it exists, to the address space of a newly
\fork-ed process. When an application crashes, \sys re-randomizes all
mapped binaries associated with the crashing process and reclaims the
previous randomized code from the cache after all processes are moved
to a newly re-randomized code. \sys prevents direct reading of
randomized code from userspace using XoM. \sys initializes a shadow
stack whenever \clone-ing a task\footnote{In this paper, a term
  \emph{task} denotes both process and thread as the convention in
  Linux kernel.}.

\subsubsection{Process Memory Layout}
\label{s:dk:mem}

\autoref{f:mem} illustrates the memory layout of two \sys processes,
\cc{websrv} (top left) and \cc{dbsrv} (top right). \sys compiler
generates a PC-relative binary with
trampoline code and fixup information \BC{1}. When the binary is first
loaded to be mapped to a process, \sys kernel first extracts
all \sys metadata--described in~\autoref{s:dc:sharing}--in the binary
and associates it on a per-file basis with the binary's \inode
structure.
After extracting metadata, \sys has the
information it needs to perform load-time randomization \BC{2}. Note
that load-time randomization and run-time re-randomization follow the
exact same procedure.
\sys first generates a random offset to set apart the code and
trampoline regions and then places functions in a random order (\ie, a
random permutation) within the code region.
Once functions are placed, \sys then uses the cached
\sys metadata to perform patching of offsets within both trampoline
and code regions, updating PC-relative \jmp targets and data locations
to preserve program semantics. Finally, the randomized code is now
semantically correct and can be cached and mapped to multiple
applications \BC{3}.

\subsubsection{Fine-Grain Code Randomization}
\label{s:dk:rand}

\PP{Allocating a virtual code region}
\sys kernel randomizes the binary when the binary is mapped with
executable permissions. For each randomized binary, the \sys kernel
allocates a 2~GB \emph{virtual} address region\footnote{
We note that, for the unused region,
we map all those virtual addresses to a single abort page
that generate a crash when accessed
to not to waste real physical memory and also detect potential attack attempts.
} (\autoref{f:mem}
\BC{2}), which will be mapped to userspace virtual address space with
coarse-grained ASLR (\autoref{f:mem} \BC{3})\footnote{We choose 2~GB
because in x86-64 architecture PC-relative addressing can refer $\pm2$~GB
range from \rip.}. \sys kernel positions the trampoline code at the
end of the virtual address region and returns the start address of the
trampoline as a result of \mmap.
The trampoline address will remain static throughout program
execution even after re-randomization.

\PP{Randomizing the code within the virtual region}
To achieve a high entropy, \sys kernel uses fine-grained randomization
within the allocated virtual address region. Once the trampoline is
positioned, \sys kernel randomly places the non-trampoline code within
the virtual address region (sized 2~GB);
\sys decides the \emph{random offset} between the code and the trampoline regions.
Once the code region is decided, \sys
permutes functions within the code region to further increase
entropy. As a result, trampoline addresses do not leak information on
non-trampoline code and an adversary cannot infer any actual codes'
location from the system information, \cc{/proc/<pid>/maps},
as they will get the same mapping information for the entire 2~GB region.

\PP{Patching the randomized code}
After permuting functions, \sys kernel patches instructions
accessing code or data according to randomization.
\sys kernel patches \rip-relative offsets in instructions. This
patching process is trivial at runtime;
\sys compiler generates fixup location information in the binary and
\sys kernel re-calculates and patches PC-relative offsets of
instructions according to the randomized function location. Note that
patching includes control flow transfer between the trampoline and
non-trampoline code and global data access (\ie, \cc{.data},
\cc{.bss}) as well as function calls to other shared libraries (\ie,
PLT/GOT).

\subsubsection{Randomized Code Cache}
\label{s:dk:cache}

\sys kernel manages a cache of randomized code. When a userspace
process tries to map a file with executable permissions, \sys kernel
first looks up whether there exists a randomized code of the file in
cache.
If cache hits, \sys kernel maps the randomized code region to the
virtual address of the requested process. Upon cache miss, it performs
load-time randomization as described earlier.
\sys kernel manages how many times the randomized code region is
mapped to userspace. If the reference counter is zero and the memory
pressure of the system is high, \sys kernel evicts the randomized
code. Thus, in normal cases without re-randomization, \sys randomizes
a binary file only once. In our implementation, the randomized code
cache is associated with the \inode cache. Thus, when the \inode is
evicted from the \inode cache under severe memory pressure, its
associated randomized code is also evicted.

\subsubsection{Execute-Only Memory (XoM)}
\label{s:dk:XoM}

We designed XoM based on Intel MPK~\cite{intel:sdm}\footnote{As of this
writing, Intel Xeon Scalable Processors~\cite{intel:xeon-scalable}
and Amazon EC2 C5 instance~\cite{amazon:c5} support MPK. Other than
the x86 architecture, ARM AArch64 architecture also supports
execute-only memory~\cite{arm:XoM}.}. With MPK, each page is assigned
to one of 16 domains (referred to as a \emph{protection key}), which
is encoded in a page table entry. Read and write permissions of each
domain can be independently controlled through an MPK register.
When randomized code is mapped to a userspace virtual address, we set
the permissions of the corresponding page table entries to
executable, which is in fact executable and readable, and
assign code memory to the XoM domain. \sys kernel configures the XoM
domain to non-accessible (\ie, neither readable nor writable) so \sys
kernel can enforce execute-only permission with MPK.
If an adversary tries to read XoM-protected code memory, \sys kernel will
raise \sigbus and trigger re-randomization.
Unlike EPT-based XoM designs~\cite{crane:readactor, tang:heisenbyte},
our MPK-based design does not impose runtime overhead.

\subsubsection{On-Demand Re-randomization}
\label{s:dk:re-rand}

\begin{figure*}[t!]
  \vspace{-.6em}
  \centering
  \includegraphics[width=\textwidth]{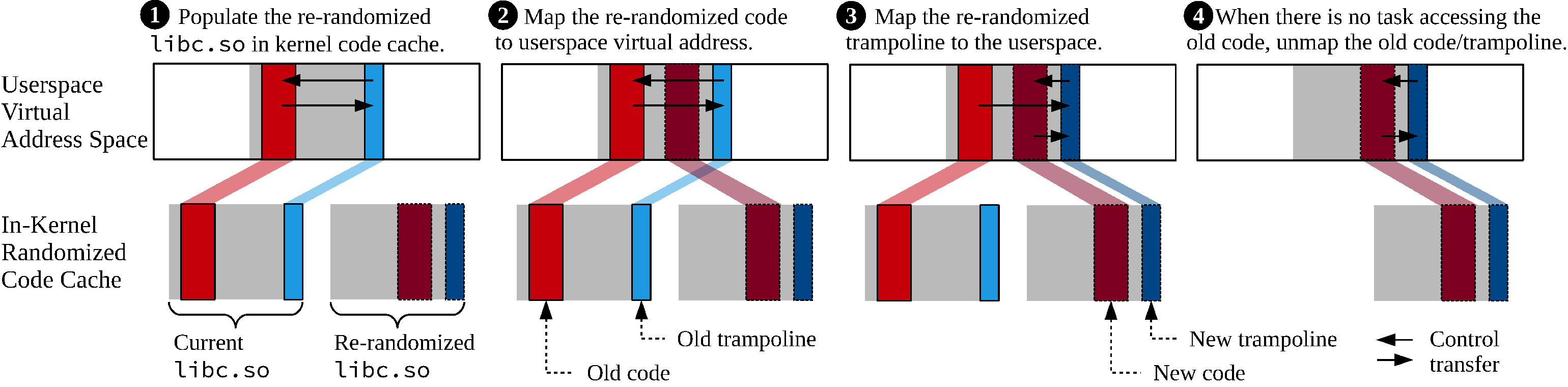}
  \caption{
    Re-randomization procedure in \sys.
    Once a new re-randomized code is populated \protect \BC{1}, \sys
    kernel maps new code and trampoline in order \protect \BC{2},
    \protect \BC{3}. This makes threads crossing the new trampoline
    migrate to the newly re-randomized code. After it is guaranteed that all threads
    are migrated to the new code, \sys reclaims the old code \protect
    \BC{4}. Unlike previous continuous per-process re-randomization
    approaches, our re-randomization is time-bound, almost zero
	overhead, and system-wide.
  }
  \spacebelowcaption
 \label{f:rerand}
\end{figure*}

\PP{Triggering re-randomization}
An unsuccessful probing of the attack causes the process to crash.
Therefore, when a process crashes \sys triggers re-randomization
of \emph{all} binaries mapped  to the crashing process.
Since \sys re-randomization thwarts attacker's knowledge (\ie, each
attempt is an independent trial), an adversary must succeed in her
first try without crashing, which is practically infeasible.

\PP{Re-randomizing code}
Upon re-randomization, \sys kernel first populates another copy of
the code (\eg, \libc) in the code cache and freshly randomizes
it (\autoref{f:rerand} \BC{1}). \sys places the trampoline code at
the same location not to change trampoline addresses to avoid mutating
code pointers but it randomly places the non-trampoline code (\ie,
random offset in \autoref{f:mem} \BC{2}) such that the new one does
not overlap with the old one. Then, it permutes functions in the
code. Thus, the re-randomized code is completely different
from the previous one without changing trampoline addresses.

\PP{Live thread migration without stopping the world}
The re-randomized code prepared in the previous step is not yet
visible to userspace processes because it is not yet mapped to
userspace virtual address space.
To make it visible,
\sys first maps the new non-trampoline code to
application's virtual address space, \autoref{f:rerand} \BC{2}. Because the old trampoline
code is still mapped, the new code is not reachable yet.
Then, \sys remaps the virtual address range of the trampoline code to
the new trampoline code by updating corresponding page table entries
\BC{3}. After this, the new trampoline code will transfer control flow
to the new non-trampoline code so that any thread crossing the
trampoline migrates to the new non-trampoline code without stopping
the world.

\PP{Safely reclaiming the old code}
\sys can safely reclaim the code
only after all threads migrates to the new code \BC{4}. \sys uses
\emph{reference counting} for each randomized code to check if there
is a thread accessing the old code.
After the new trampoline code is mapped \BC{3}, \sys sets a reference
counter of the old code to the number of all \emph{runnable}
tasks~\footnote{A task in a \cc{TASK_RUNNING} status in Linux kernel.}
that map the old code. It is not necessary to wait for migration of
non-runnable, sleeping task because it will correctly migrate to the
newest randomized code region when it passes through the (virtually)
static return trampoline, which refers to the new layout when it wakes
up.
The reference counter is decremented when a runnable task
enters into \sys kernel due to system call or preemption.
When calling a system call, \sys kernel will decrement reference
counters of all code that needs to be reclaimed. When the task returns
to userspace, it will return to the return trampoline and the return
trampoline will transfer to the new code.
When a task is preempted out, it may be in the middle of executing the
old non-trampoline code. Thus, \sys kernel not only decrements
reference counters but also translates \rip of the task to the
corresponding address in the new code. Since \sys permutes at function
granularity, \rip translation is merely adding an offset between the
old and new function locations.

\PP{Summary}
Our re-randomization scheme has three nice properties: time boundness
of re-randomization, almost zero overhead of running process, and
system-wide re-randomization.
The re-randomization is guaranteed to finish at most within one
scheduling quantum (\eg, 1 msec) once the newly randomized code is
exposed \BC{3}. That is because \sys migrates \emph{runnable} tasks at
system call and scheduling boundary.
If another process crashes in the middle of re-randomization, \sys
will not trigger another re-randomization until the current
randomization finishes. However, as soon as the new randomized code is
populated \BC{1}, a new process will map the new code
immediately. Therefore, the old code cannot be observed more than once.
\sys kernel populates a new randomized code in the context of a
crashing process. All other runnable tasks only additionally perform
reference counting or translation of \rip to the new code. Thus, its
runtime overhead for runnable tasks is negligible.
\emph{To the best of our knowledge, \sys is the first system to
perform system-wide re-randomization allowing code sharing.}

\subsubsection{Shadow Stack}
\label{s:dk:thread}
To hide the shadow stack location, \sys first reserves a 2~GB of virtual
memory space with abort pages and then chooses a base address
to map the shadow stack.
Additionally,
no direct reference to the shadow stack is available in memory because
\sys accesses it via a dedicated register, \gs,
which will not disclose the base address of the shadow stack.
As a result, an adversary must
brute-forcingly guess its base address;
any crash in such attempt will trigger
re-randomization, which invalidates all prior information gained.
When a new task is created (\clone),
the \sys kernel allocates a new shadow stack and
copies parent's shadow stack to its child.

\section{Implementation}
\label{s:impl}

We implemented \sys on the Linux x86-64 platform. \sys compiler is
implemented using LLVM 6.0.0 and \sys kernel is implemented based on
Linux kernel 4.17.0 modifying \totalModifiedLLVM~ and
\totalModifiedKernel~ lines of code (LOC), respectively.
We used \cc{musl libc} 1.1.20~\cite{musl-web}, which is a fast,
lightweight C standard library implementation for Linux, because
\cc{glibc} cannot be compiled with Clang. We manually modified
\totalModifiedMusl~ LOC in \cc{musl libc} to make assembly functions
(\eg, math functions and atomic intrinsics) follow the \sys calling
convention by adding C wrapper functions so \sys compiler can
automatically identify and instrument them.

\subsection{\sys Compiler}
\label{s:impl:compiler}

\PP{Trampoline}
\sys compiler is implemented as backend target-ISA (x86) specific
\machinefunctionpass. This pass instruments each function body as
described in \autoref{s:design-compiler}.

\PP{Re-randomizable code}
To force all instructions to use PC-relative addressing, \sys compiler
uses \cc{-fPIC}.
As an optimization of our trampoline design, \sys
compiler uses a register \rbp for stack top index of a shadow
stack. To force the compiler to relinquish the use of the register
\rbp, \sys compiler uses \cc{-fomit-frame-pointer}.
To save space in PC-relative addressing of code and data, the x86
architecture provides various \jmp instruction variants varying in
offset size from 1, 2, or 4 bytes. In \sys, to maximize entropy and be
able to use the full span of memory within our declared 2 GB virtual
address region, \sys compiler uses \cc{-mrelax-all} to force the
compiler to always emit full 4-byte displacement in the
executable. Since all PC-relative instructions have 4-byte wide
displacement, \sys kernel can freely place any function to within 2 GB
address range without any restriction.
To completely separate code and data for XoM, \sys compiler disables
jump tables in \cc{.text} section using \cc{-fno-jump-tables}.

\subsection{\sys Kernel}
\label{s:impl:kernel}

\PP{Shadow stack}
Currently the maximum shadow stack size is set to 64~KB;
when a task is created,
\sys kernel creates a 2~GB virtual address space region
and randomly place its shadow stack in that region,
guarded by inaccessible pages (to hide the shadow stack).
To maximize performance, \sys implements a compact shadow
stack without comparisons~\cite{shadesmar-burow-oakland19}.

\PP{Secure random number generation}
To perform randomization,
\sys uses cryptographically secure random number generator
in Linux kernel based on hardware random sources such as
\rdrand/\rdseed instructions in modern Intel architectures.

\subsection{Limitation of Our Prototype}
\label{s:impl:limitation}

\PP{Assembly Code}
\sys does not support inline assembly as
it difficult to figure how to deal with module-level inline assembly
as was present in \cc{musl libc}; however, this could be resolved with further engineering.
Instead, we manually added C
wrapper functions so \sys compiler adds trampolines which complies to
\sys calling convention.

\PP{Setjmp and exception handling}
\sys uses a shadow stack to store return addresses. Thus,
functions such as \cc{setjmp}, \cc{longjmp}, and \cc{libunwind} that
directly manipulate return addresses on stack are not supported by our
current prototype. However, modifying these functions
is straightforward because essentially our shadow
stack is a variant of compact, register-based shadow
stack~\cite{shadesmar-burow-oakland19}.

\PP{C++ support}
Our prototype does not support C++ applications since we do not have
a stable standard C++ library that is \cc{musl}-compatible. Therefore
handling C++ exceptions and protecting vtables is out of scope.
\section{Evaluation}
\label{s:eval}
We evaluate \sys by answering these questions:

\squishlists
\item How secure is \sys, when presented against current known attacks
  on randomization?
 (\autoref{s:eval:sec})

\item How much performance overhead does \sys impose, particularly
  for compute-intensive benchmarks?
 (\autoref{s:eval:perf})

\item How scalable is \sys in a real-world network facing server,
  particularly
  with concurrent processes,
  in terms of load time,
  re-randomization time, and memory used?
  (\autoref{s:eval:scal})
\squishend

\PP{Applications}
We evaluate the performance overhead and scalability of \sys using SPEC CPU2006
and NGINX web server.
The SPEC benchmark suite has various realistic compute-intensive applications
(\eg, \cc{gcc}) which are ideal to see the worst-case performance overhead of
\sys.
We tested all 12 C language benchmarks; we excluded C++ benchmarks as our
current prototype does not support it.
We choose SPEC CPU2006 over its newer version, SPEC CPU2017, because SPEC
CPU2006 has been popularly used to show the performance overhead in many prior
works.
Input size \emph{ref} was used for all benchmarks.
To test performance and scalability of \sys on a complex, real-world
application, we ran NGINX, which is a widely utilized web server. We
configured NGINX as multi-process.
We report the average of four runs.

\PP{Experimental setup}
All programs are compiled with optimization \cc{-O2} and
run on a 24-core (48-hardware threads) machine equipped with two Intel
Xeon Silver 4116 CPUs (2.10~GHz) and 128~GB DRAM.

\subsection{Security Evaluation}
\label{s:eval:sec}

To evaluate the security of \sys, we first analyze the resiliency of
\sys against existing attacker models against load-time randomization
(A1--A2, \autoref{s:eval:sec:1r}) and continuous re-randomization (A3--A4,
\autoref{s:eval:sec:rr}).
Then, to illustrate the effectiveness of \sys
for a wider class of code-reuse attacks beyond ROP,
we present results of the residual attack surface analysis
using the threat model of NEWTON~\cite{van:newton}
(\autoref{s:eval:sec:beyondrop}).

\boxbeg
{\bf \sys Security Summary:}
\small
\squishlists
\item \hspace{-0.3em}{\it vs.}\hspace{0.1em}{\bf A1:} Execute-only memory blocks the attack.
\item \hspace{-0.3em}{\it vs.}\hspace{0.1em}{\bf A2:} Re-randomization blocks any code inference via crash.
\item \hspace{-0.3em}{\it vs.}\hspace{0.1em}{\bf A3:}
	Execute-only memory and a large search space (2~GB dummy mappings)
	block JIT-ROP and crash-resistant probing.
\item \hspace{-0.3em}{\it vs.}\hspace{0.1em}{\bf A4:} Trampolines decouple function entry from function bodies blocking any type of code pointer offsetting; full function code reuse of exported functions remains possible.
\item \hspace{-0.3em}{\it vs.}\hspace{0.1em}{\bf NEWTON:} the same as A4.
\squishend
\boxend

\subsubsection{Attacks against Load-Time Randomization}
\label{s:eval:sec:1r}

%%%%%%%%%%%%
\PP{Against JIT-ROP attacks (A1)}
\sys asserts permissions for all code areas
(both code and trampoline regions) as execute-only (via XoM);
thereby, an attacker with JIT-ROP
capability cannot read code contents directly.

%%%%%%%%%%%%
\PP{Against code inference attacks (A2)}
\sys blocks code inference attacks, including BROP~\cite{bittau:brop},
clone-probing~\cite{lu:runtimeASLR},
and destructive code read
attacks~\cite{zombie-gadget-snow-oakland16, bgdx-pewny-acsac17} via
layout re-randomization triggered by an application crash or XoM violation.
This mechanism effectively blocks A2 attacks by
preventing attackers from accumulating indirect information
because every re-randomization renders
all previously gathered (if any) information regarding
the code layout invalid.

%%%%%%%%%%%%
\PP{Hiding shadow stack}
Attackers with arbitrary read/write capability (A1/A2)
may attempt to leak/alter shadow stack contents if
its address is known.
Although the location of the shadow stack is hidden behind
the \cc{\%gs} register to prevent leakage of pointers,
attackers may employ attacks that undermine this
sparse-memory based information
hiding~\cite{oikonomopoulos:pokingcpi,
  goktas:information-hiding,evans:missingcpi}.
To prevent such attacks,
\sys reserves a 2~GB virtual memory space for the shadow stack
(the same way \sys allocates code/library space)
and then chooses a random offset to map the shadow stack;
other pages in the remaining 2~GB space are mapped as
an abort page that has no permissions.
Even assuming if an attacker is able to identify
the 2~GB region for the shadow stack using crash-less
poking~\cite{goktas:information-hiding} or employing allocation
oracles~\cite{oikonomopoulos:pokingcpi},
they must also overcome the randomization entropy
of the offset to get a valid address within this region;
any incorrect probe will generate crash (due to abort pages),
thereby, thwarting the attack.
Consequently,
the probability of successfully guessing the location of
any valid shadow stack address is roughly
one in $2^{31}$, practically infeasible.

%%%%%%%%%%%%
\PP{Entropy}
\sys applies both function-level permutation and
random start offset to provide high entropy to
the new code layout.
Specifically,
\sys permutes all functions in each executable and
applies a random start offset to
the code area in 2~GB space for each randomization.
Thus, randomization entropy depends on the number of functions
in the executable and the size of a code region
(\ie, $log_2(n! \cdot 2^{31})$ where $n$ is the number of functions).
To give an idea of how much entropy \sys provides,
we take an example of \cc{470.lbm} in SPEC CPU2006,
a case which provides the minimum entropy in our evaluation.
The program, which contains 26 functions and
is less than 64~KB in size,
has 119.38 bits entropy.
Therefore,
even for a small program,
\sys randomizes the code with significantly high entropy (119 bits) to
render attacker's success rate for guessing the layout negligible.

\subsubsection{Attacks against Continuous Re-randomization}
\label{s:eval:sec:rr}

%%%%%%%%%%%%
\PP{Against low-profile attacks (A3)}
\sys does not rely on timing nor system call history for
triggering re-randomization.
As a result,
neither low-latency attacks
nor attacks without involving system calls are
effective against \sys.
Instead,
re-randomization is triggered and performed by
any \sys instrumented application process on the system that
encounters a crash (\eg, XoM violation).
Nonetheless,
a potential A3 vector could be one that
does not cause any crash during exploitation
(\eg, attackers may employ
crash-resistant probing~\cite{oikonomopoulos:pokingcpi,
  goktas:information-hiding, evans:missingcpi, gawlik:crash-resist,
  kollenda:auto-crash-resist}).
In this regard, \sys places all code in execute-only memory with
2~GB mapped region.
Such a stealth attack could only identify multiples of 2~GB code
regions and will fail to leak any fine-grained layout of code or
code addresses stored in trampolines.

%%%%%%%%%%%%
\PP{Against code pointer offsetting attacks (A4)}
For code pointers referring to call trampolines,
attackers may attempt to launch an A4 attack by
adding/subtracting an offset to the pointer.
To defend against such an attack,
\sys decouples any correlation between trampoline function \emph{entry} addresses and
function \emph{body} addresses (\ie, no fixed offset),
so attackers cannot refer to the middle of a function for
a ROP gadget without actually obtaining a valid function body address.
Additionally,
the trampoline region is also protected with XoM,
thus attackers cannot probe it to obtain
function body addresses to launch A4.
\sys limits available code-reuse targets to
only exported functions in the trampoline region.
We analyze the residual attack surface of \sys in
\autoref{tbl:att-surface}.

\subsubsection{Viable Attacks in \sys}
\label{s:eval:sec:beyondrop}

\densetbl
\begin{table}[t]
  \vspace{-.4em}
  \ra{1}
  \centering
  \ssmall
  \newcommand{\rowI}{perlbench}
\newcommand{\rowII}{bzip2}
\newcommand{\rowIII}{gcc}
\newcommand{\rowIV}{mcf}
\newcommand{\rowV}{milc}
\newcommand{\rowVI}{gobmk}
\newcommand{\rowVII}{hmmer}
\newcommand{\rowVIII}{sjeng}
\newcommand{\rowIX}{libquantum}
\newcommand{\rowX}{h264ref}
\newcommand{\rowXI}{lbm}
\newcommand{\rowXII}{sphinx3}
\newcommand{\rowXIII}{NGINX}
\newcommand{\rowXIV}{musl libc}
\newcommand{\rowXV}{Total}

\begin{tabular}{l|r|r|r|r|r}
  \toprule

  % first line
  \multirow{2}{*}{\textbf{Benchmark}} &
  \multicolumn{2}{c|}{\textbf{Return address}} &
  \textbf{ROP gadget} &
  \textbf{Func reuse} &
  \textbf{\sys}
  \\
  % The following specify horizontal divider lines
  \cmidrule(r){2-3}
  \cmidrule(r){4-4}
  \cmidrule(r){4-4}
  \cmidrule(r){5-5}

  % second line
  &
  \textbf{\# \cc{call}} &
  \textbf{\# \cc{i-call}} &
  \textbf{\# \cc{gadgets}} &
  \textbf{\# fn entry} &
  \textbf{reduction (\%)}
  \\
  \midrule
	\textbf{\rowI}  &13963  &272    &34371  &1668   & 48606/50274 (96.7\%) \\
	\textbf{\rowII} &277	&56     &1569   &81     & 1902/1983 (95.9\%) \\
	\textbf{\rowIII} &48096 &518    &89746  &4318   & 138360/142678 (97.0\%) \\
	\textbf{\rowIV} &77     &3      &513    &33     & 593/626 (94.7\%) \\
	\textbf{\rowV} &2104    &7      &3864   &244    & 5975/6219 (96.1\%) \\
	\textbf{\rowVI} &9521   &48     &37999  &2477   & 47568/50045 (95.0\%) \\
	\textbf{\rowVII} &4237  &12     &9466   &478    & 13715/14193 (96.6\%) \\
	\textbf{\rowVIII} &1054 &4      &3110   &139    & 4168/4307 (96.8\%) \\
	\textbf{\rowIX} &469    &3      &1686   &107    & 2158/2265 (95.3\%) \\
	\textbf{\rowX} &3118    &372    &14456  &528    & 17946/18474 (97.1\%) \\
	\textbf{\rowXI} &70     &3      &394    &26     & 467/493 (94.7\%) \\
	\textbf{\rowXII} &2714  &11     &5628   &326    & 8353/8679 (96.2\%) \\
	\textbf{\rowXIII} &5316 &309    &15434  &1565   & 21059/22624 (93.1\%) \\
	\textbf{\rowXIV} &4400  &77     &29743  &3722   & 34220/37942 (90.2\%) \\
  \midrule
	\textbf{\rowXV} &95416  &1695   &247979 &15712  & 345090/360802 (95.6\%) \\
  \bottomrule
\end{tabular}

  \caption{
    Potential attack surface and \sys's reduction in binaries.
    While a function call trampoline is a only reusable target in
    \sys, an attacker cannot infer ROP gadgets from the call
    trampoline addresses.
  }
  \spacebelowcaption
  \vspace{-.8em}
  \label{tbl:att-surface}
\end{table}
\densetblend

%%%%%%%%%%%%
\PP{Attack analysis with NEWTON}
To measure the boundary of
viable attacks against \sys,
we present a security analysis of \sys based on
the threat model set by NEWTON~\cite{van:newton}.
In this regard,
we analyze possible writable pointers that
can change the control flow of a program (write constraints)
as well as
possible available gadgets in \sys (target constraints),
which will reveal what attackers can do under this threat model.
In short,
\sys allows
only the reuse of
exported functions via call trampolines.

For write constraints,
attackers cannot overwrite real code addresses such as
return addresses and code addresses in the trampoline.
\sys only allows attackers to overwrite
other types of pointer memory, \eg,
object pointers and pointers to the call trampoline.
For target constraints,
attackers can reuse only the exported functions via call trampoline.
Note that a function pointer is a reusable target in any
re-randomization techniques using immutable code
pointers~\cite{wang:reranz, williams:shuffler, chen:codearmor}.
Although \sys allows attackers to reuse
function pointers
in accessible memory (\eg, a function pointer in a structure), such
live addresses will never include real code addresses, such as a
return address or real code address, and will be limited to addresses
referencing call trampolines.
\emph{Under these write and target constraints, inferring the location
of ROP gadgets from code pointers (\eg, leaking code addresses or
adding an offset) is not possible.}

\PP{Residual attack surface}
\autoref{tbl:att-surface} presents
potential code-reuse attack surface of
programs in this evaluation and
how much \sys reduces such attack surface
(\ie, possible code reuse targets in the program).
For each program,
we present the attack surface in three categories.
\squishlists
\item Direct calls (\cc{\# call}) and indirect calls (\cc{\# i-call}),
which may leak return addresses to the regular stack.
\item ROP Gadgets (\cc{\# gadgets}),
the main ingredient in constructing a code re-use payload
by being chained together
to make a valid attack.
This data is obtained via running ROPgadget~\cite{ROPGadget} on each
benchmark binary.
\item Function entries (\cc{\# fn entry}) that can be used for whole
function re-use attack.
\squishend

\sys completely protects the first two categories
(\ie, return address and ROP gadget)
using shadow stack and XoM, respectively.
Therefore,
the remaining potential attack surface is function entry
(\ie, call trampolines in \sys).
Evaluating SPEC, \cc{musl libc}, and NGINX,
\sys reduces access to these sensitive fragments up to
a max of 97.1\% and 95.6\% on average,
constraining attackers to reuse only exported function entries.
\subsection{Performance Evaluation}
\label{s:eval:perf}

\begin{figure}
  \vspace{-.8em}
  \centering
  \includegraphics[width=\columnwidth]{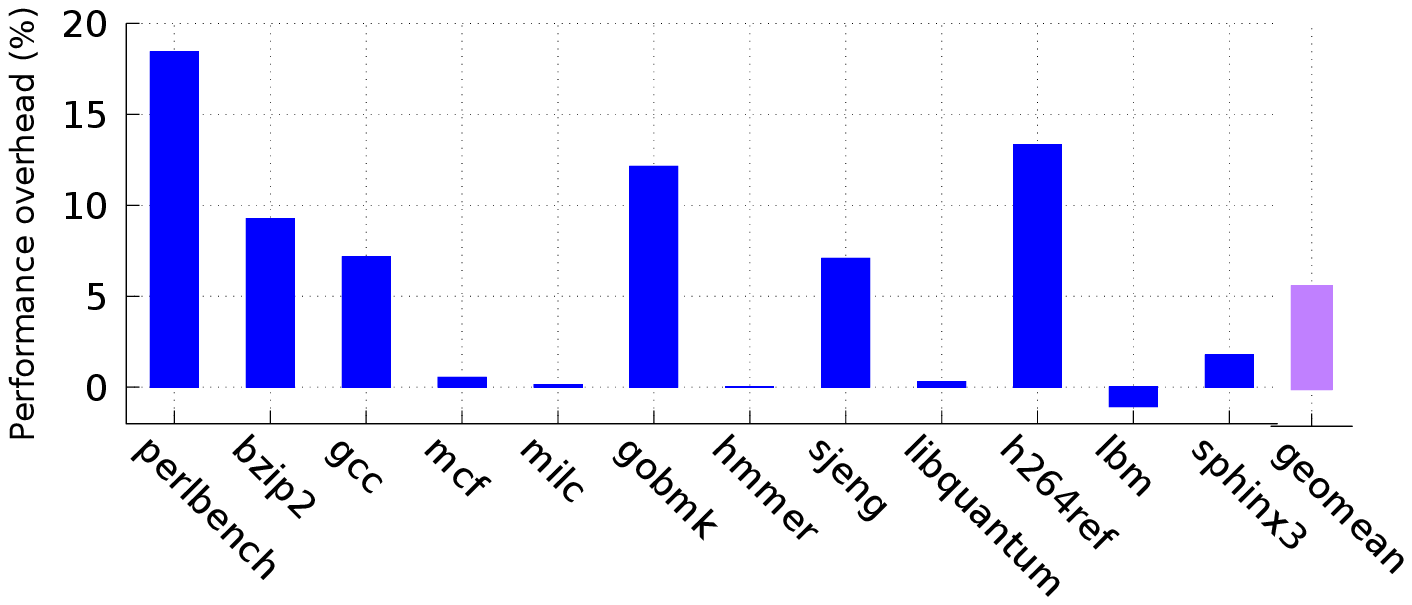}
  \vspace{-2.7em}
  \caption{Performance overhead of \sys for SPEC CPU2006}
  \spacebelowcaption
  \label{f:overhead}
  \vspace{2.2em}
  \centering
  \includegraphics[width=\columnwidth]{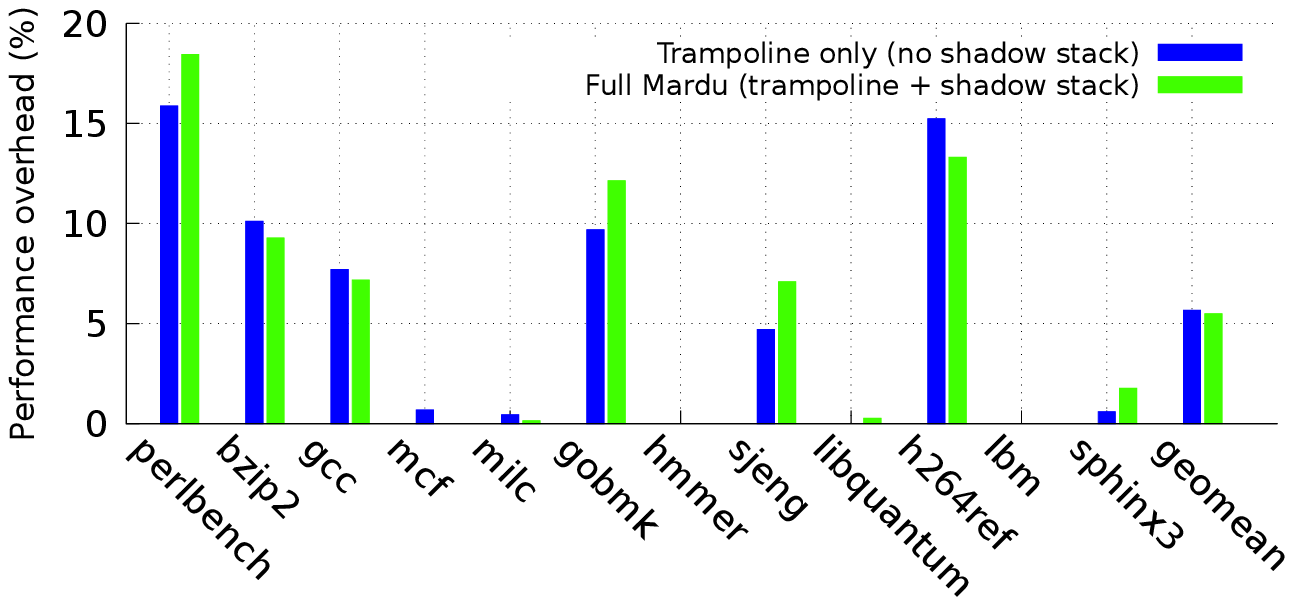}
  \vspace{-2.7em}
  \caption{Overhead breakdown for SPEC CPU2006}
  \spacebelowcaption
  \label{f:breakdown}
\end{figure}

\PP{Runtime performance overhead with SPEC CPU2006}
\autoref{f:overhead} shows the relative performance overhead of
SPEC with \sys compared to the unprotected baseline,
compiled with vanilla Clang.
Overall, \sys's average overhead is comparable to the fastest
re-randomization systems.
Notably, \sys worst-case overhead is
significantly better than similar systems.
The average overhead of \sys is 5.5\%, and the worst-case overhead is
18.3\% (\cc{perlbench});
in comparison to Shuffler~\cite{williams:shuffler} and
CodeArmor~\cite{chen:codearmor}, whose reported average overheads are
14.9\% and 3.2\%, respectively, while their worst-case overhead are 45\%
and 55\%, respectively (see \autoref{tbl:comparison}).
This confirms \sys is capable of matching if not slightly improving
the performance (especially worst-case) overhead, while casting a
wider net in terms of known attack coverage, compared to
the current state-of-the-art.

\PP{Performance overhead breakdown}
The two prominent sources of runtime overhead in \sys are trampolines
and the shadow stack.
To understand how much runtime overhead each code transformation
imposes, we ran SPEC with two different configurations:
we first enabled only trampolines and do not use a shadow stack
(trampoline only in \autoref{f:breakdown}), then we enabled both
trampolines and the shadow stack (full \sys in \autoref{f:breakdown}).
The performance overhead of both are normalized to vanilla SPEC
compiled with vanilla Clang.
As \autoref{f:breakdown} shows, the major source of overhead is
trampoline; trampolines incur 5.9\% overhead on average, with the
worst cases being 15.8\% and 14.2\% of overhead, for \cc{perlbench}
and \cc{h264ref}, respectively.
Note that \sys's shadow stack overhead is negligible. The average
difference comparing \sys trampolines with full \sys is less than
0.3\%, and in the noticeable gaps, adding less than 2\% to \sys in
most cases compared to using \sys trampolines only.
This is expected as \sys uses a compact shadow stack without
comparison epilogue. This implementation performs only essential
bookkeeping to utilize the shadow stack; skipping unnecessary
epilogue micro-optimizations~\cite{shadesmar-burow-oakland19}.

\subsection{Scalability Evaluation}
\label{s:eval:scal}

\begin{figure}
  \vspace{-.8em}
  \centering
  \includegraphics[width=\columnwidth]{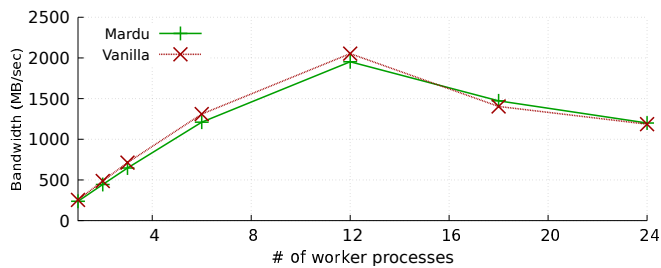}
  \vspace{-.3em}
  \caption{Performance comparison of NGINX web server}
  \spacebelowcaption
  \label{f:nginx}
\end{figure}

\PP{Runtime performance overhead with NGINX}
NGINX is configured to accommodate a maximum of 1024 connections per
processor, and its performance is observed according to the
number of worker processes. \cc{wrk}~\cite{github:wrk} is used to
generate HTTP requests for benchmarking. \cc{wrk} spawns the same
number of threads as NGINX workers and each \cc{wrk} thread sends
a request for a 6745-byte static html. To see the worst-case
performance, \cc{wrk} is run on the same machine as NGINX to factor
out network latency.
\autoref{f:nginx} presents the performance of NGINX with and without
\sys for a varying number of worker processes. The performance
observed shows that \sys exhibits very similar throughput to
vanilla. \sys incurs 4.4\%, 4.8\%, and 1.2\% throughput degradation on
average, at peak (12 threads), and at saturation (24 threads),
respectively.
Note that Shuffler~\cite{williams:shuffler} suffers from the overhead
from \emph{per-process} shuffling thread. Even in their NGINX
experiments \emph{with network latency} (\ie, running a benchmarking
client on a different machine), Shuffler shows 15-55\% slowdown.
This verifies \sys's design that having the crashing process perform
system-wide re-randomization, rather than a per-process background
thread as in Shuffler, scales better.

\begin{figure}
  \vspace{-1.em}
  \centering
  \includegraphics[width=\columnwidth]{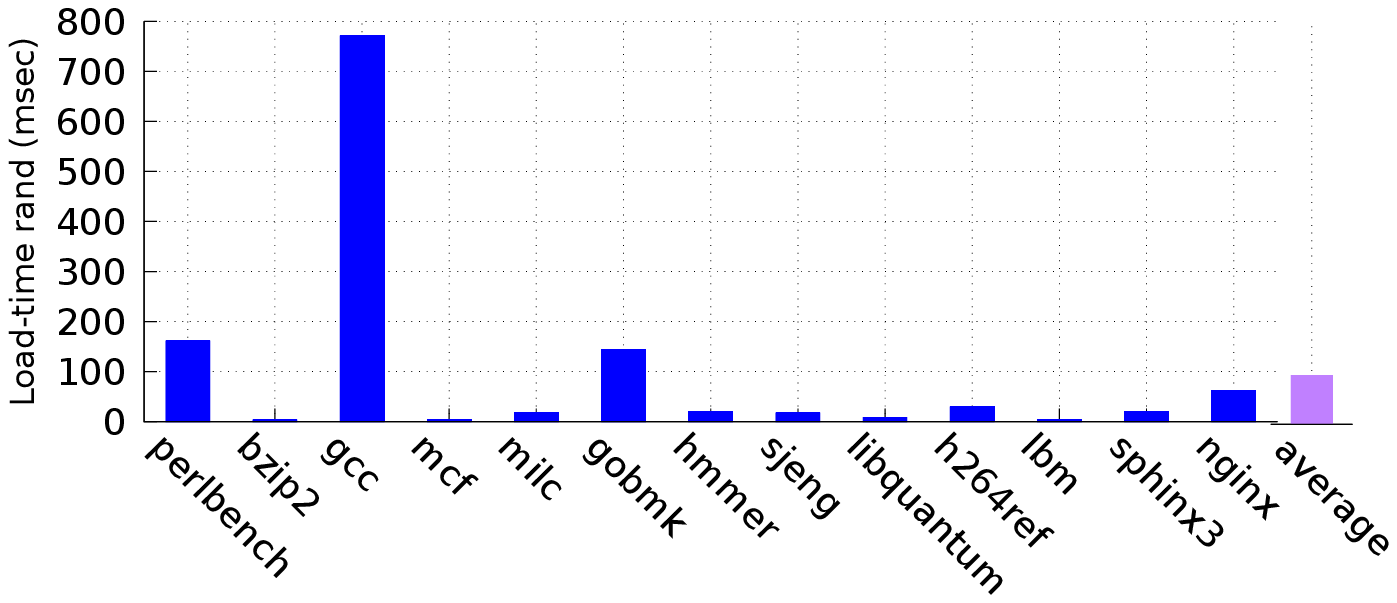}
  \vspace{-2.7em}
  \caption{Cold load-time randomization overhead}
  \spacebelowcaption
  \label{f:loadtime}
  \vspace{1.2em}
  \centering
  \includegraphics[width=\columnwidth]{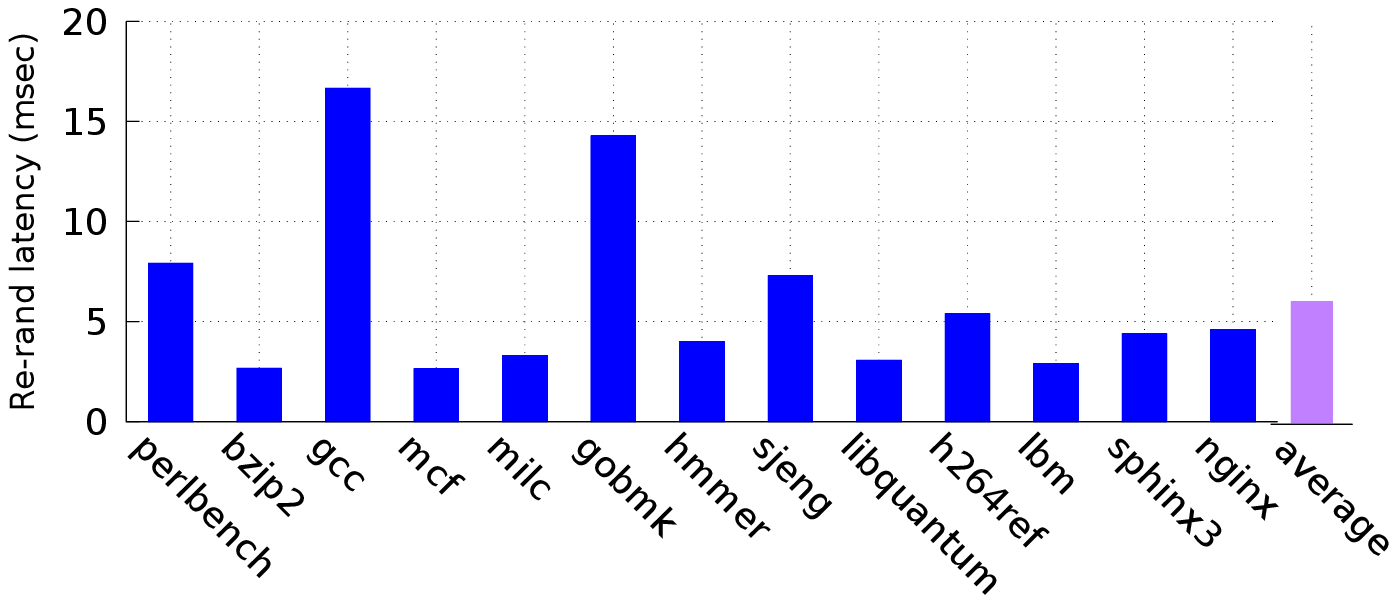}
  \vspace{-2.7em}
  \caption{Re-randomization latency}
  \spacebelowcaption
  \label{f:rerand-lat}
\end{figure}

\PP{Load-time randomization overhead}
We categorize load-time to cold or warm load-time whether
the in-kernel code cache (\BC{2} in \autoref{f:mem}) hits or not.
Upon a code cache miss (\ie, the executable is first loaded in a
system), \sys performs initial randomization including function-level
permutation, start offset randomization of the code layout, and
loading \& patching of fixup metadata. As \autoref{f:loadtime} shows,
all C SPEC benchmarks showed negligible overhead averaging 95.9
msec. \cc{gcc}, being the worst-case, takes 771 msec; it requires the most
overall fixups relative to other benchmarks (see \autoref{tbl:code-stat}). For NGINX, we
observe that load time is constant (61 msec) for any number of
specified worker processes. Cold load-time is roughly linear to the
number of trampolines in \autoref{tbl:code-stat}.
Upon a code cache hit, \sys simply maps the already-randomized code
to a user-process's virtual address space. Therefore we found that warm
load-time is negligible. Note that, for a cold load-time of \cc{musl
libc} takes about 52 msec on average. Even so, this is a one time
cost; all subsequent warm load-time accesses of fetching \cc{musl
libc} takes below 1 $\mu$sec, for any program needing it.
Thus, load time can be largely ignored.

\PP{Re-randomization latency}
\autoref{f:rerand-lat} presents the time taken to re-randomize all
associated binaries of a crashing process.
The time includes creating re-randomizing the code layout, and reclaiming the
old code (\BC{1}-\BC{4} in \autoref{f:rerand}).
To emulate an XoM violation, we killed the process with a \sigbus signal and
measured the re-randomization time inside the kernel.
The average latency of SPEC is 6.2~msec.
The difference between load-time and
re-randomization latency because \sys takes advantage of
the metadata being cached from load-time, this means
no redundant file I/O penalty is incurred, giving this performance gain.
To evaluate the efficiency of re-randomization on
multi-thread/multi-process applications, we measured the
re-randomization latency with varying number of NGINX worker processes
up to 24. We confirm that the latency is consistent regardless of
number of workers (5.8~msec on average with 0.5~msec of standard
deviation).

\begin{figure}
  \centering
  \includegraphics[width=\columnwidth]{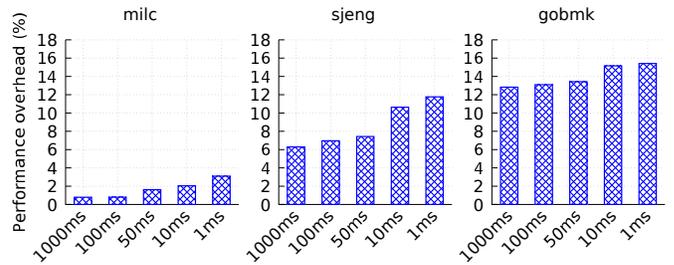}
  \vspace{-1.5em}
  \caption{Overhead varying re-randomization frequency}
  %\spacebelowcaption
  \label{f:bruteforce-rerand}
\end{figure}

\PP{Re-randomization overhead under active attacks}
A good re-randomization system should exhibit good performance not
only in its idle state but also under stress from active attacks.
To evaluate this, we stress test \sys under frequent re-randomization
to see how well it can perform, assuming a scenario that \sys is under
attack.
In particular, we measure the performance of SPEC benchmarks
while triggering frequent re-randomization. We emulate the attack by
running a background application, which continuously crashes at the
given periods: 1~sec, 100~msec, 50~msec, 10~msec, and 1~msec.
SPEC benchmarks and the crashing application are linked with the \sys
version of \cc{musl libc}, forcing \sys to constantly re-randomize
\cc{musl libc} and potentially incur performance degradation
on other processes using the same shared library.
In this experiment, we choose three representative benchmarks,
\cc{milc}, \cc{sjeng}, and \cc{gobmk}, that \sys exhibits a small,
medium, and large overhead in an idle state, respectively.
\autoref{f:bruteforce-rerand} shows that the overhead is consistent,
and in fact, is very close to the performance overhead in the idle
state observed in \autoref{f:overhead}. More specifically, all three
benchmarks differ by less than 0.4\% at a 1~sec re-randomization
interval. When we decrease the re-randomization period to 10~msec and
1~msec, the overhead is quickly saturated. Even at 1~msec re-randomization
frequency, the additional overhead is under 6~\%.
These results confirm that \sys provides performant
system-wide re-randomization even under active attack.

\PP{Runtime memory savings}
While an upfront one-time cost is paid for instrumenting with \sys,
the savings greatly outweigh this.
To illustrate, we show a typical use case of \sys in regards to shared code.
\cc{musl libc} is $\approx$800~KB in size, instrumented is 2~MB.
Specifically, \cc{musl libc} has 14K trampolines and
7.6K fixups for PC-relative addressing, the total trampoline size
is 190~KB and the amount of loaded metadata is 1.2~MB (refer to \autoref{tbl:code-stat}).
Since \sys supports code sharing, only one copy of \cc{libc} is needed for
the entire system.
Our experimental setup at idle reported 310 processes while a typical
2-core consumer laptop at idle reported 263 processes.
Therefore a rough estimate of memory savings for \cc{libc} that \sys
provides compared to similar instrumentation that did not support code
sharing is over $\approx$526-620~MB of still usable runtime memory.
Furthermore, comparing to time-based continuous re-randomization
techniques such as Shuffler~\cite{williams:shuffler} and
CodeArmor~\cite{chen:codearmor} which almost always maintain two
copies of code, \sys's memory saving for \cc{libc} is
$\approx$1-1.2~GB.
Backes~\etal~\cite{backes:oxymoron} and
Ward~\etal~\cite{relrop-wardleakage-esorics19} also highlighted the code
sharing problem in randomization techniques and reported a similar amount of
memory savings by sharing randomized code.
Finally, note that the use of shadow stack does not
increase runtime memory footprint because \sys solely relocates return
address from the normal stack to the shadow stack.

\densetbl
\begin{table}[t]
  \ra{1}
  \centering
  \ssmall
  % Variable declarations
\newcommand{\rIhI}{Benchmark}
\newcommand{\rIhII}{Numbers of Fixups}
\newcommand{\rIhIII}{Binary Increase (bytes)}

\newcommand{\rIIhI}{Call Tr.}
\newcommand{\rIIhII}{Ret Tr.}
\newcommand{\rIIhIII}{PC-rel. addr}
\newcommand{\rIIhIV}{Total}
\newcommand{\rIIhV}{Trampolines}
\newcommand{\rIIhVI}{Metadata}
\newcommand{\rIIhVII}{Total}

\newcommand{\rowI}{perlbench}
\newcommand{\rowII}{bzip2}
\newcommand{\rowIII}{gcc}
\newcommand{\rowIV}{mcf}
\newcommand{\rowV}{milc}
\newcommand{\rowVI}{gobmk}
\newcommand{\rowVII}{hmmer}
\newcommand{\rowVIII}{sjeng}
\newcommand{\rowIX}{libquantum}
\newcommand{\rowX}{h264ref}
\newcommand{\rowXI}{lbm}
\newcommand{\rowXII}{sphinx3}
\newcommand{\rowXIII}{NGINX}
\newcommand{\rowXIV}{musl libc}

\begin{tabular}{l|rrrr|rrr} 
  \toprule
  
  \multirow{2}{*}{\textbf{\rIhI}} &
  \multicolumn{4}{c|}{\textbf{\rIhII}} & 
  \multicolumn{3}{c}{\textbf{\rIhIII}}
  \\
  \cmidrule(r){2-8} &
  
  % First Group: Number of fixups
  {\textbf{\rIIhI}} & {\textbf{\rIIhII}} & {\textbf{\rIIhIII}} & {\textbf{\rIIhIV}} &
  % Second Group: Binary Increase
  {\textbf{\rIIhV}} & {\textbf{\rIIhVI}} & {\textbf{\rIIhVII}} \\
  \midrule
\textbf{\rowI} & 
	\the\numexpr\ENTRYperlbench\relax &
	\the\numexpr\RETperlbench\relax &
	\the\numexpr\PATCHperlbench\relax &
	\the\numexpr\TOTALperlbench\relax &
	\the\numexpr\RESperlbench\relax &
	\the\numexpr\METAperlbench\relax &
	\the\numexpr\TOTALIIperlbench\relax \\
\textbf{\rowII} &
	\the\numexpr\ENTRYbzip\relax &
	\the\numexpr\RETbzip\relax &
	\the\numexpr\PATCHbzip\relax &
	\the\numexpr\TOTALbzip\relax &
	\the\numexpr\RESbzip\relax &
	\the\numexpr\METAbzip\relax &
	\the\numexpr\TOTALIIbzip\relax \\
\textbf{\rowIII} &
	\the\numexpr\ENTRYgcc\relax &
	\the\numexpr\RETgcc\relax &
	\the\numexpr\PATCHgcc\relax &
	\the\numexpr\TOTALgcc\relax &
	\the\numexpr\RESgcc\relax &
	\the\numexpr\METAgcc\relax &
	\the\numexpr\TOTALIIgcc\relax \\
\textbf{\rowIV} &
	\the\numexpr\ENTRYmcf\relax &
	\the\numexpr\RETmcf\relax &
	\the\numexpr\PATCHmcf\relax &
	\the\numexpr\TOTALmcf\relax &
	\the\numexpr\RESmcf\relax &
	\the\numexpr\METAmcf\relax &
	\the\numexpr\TOTALIImcf\relax \\
\textbf{\rowV} &
	\the\numexpr\ENTRYmilc\relax &
	\the\numexpr\RETmilc\relax &
	\the\numexpr\PATCHmilc\relax &
	\the\numexpr\TOTALmilc\relax &
	\the\numexpr\RESmilc\relax &
	\the\numexpr\METAmilc\relax &
	\the\numexpr\TOTALIImilc\relax \\
\textbf{\rowVI} &
	\the\numexpr\ENTRYgobmk\relax &
	\the\numexpr\RETgobmk\relax &
	\the\numexpr\PATCHgobmk\relax &
	\the\numexpr\TOTALgobmk\relax &
	\the\numexpr\RESgobmk\relax &
	\the\numexpr\METAgobmk\relax &
	\the\numexpr\TOTALIIgobmk\relax \\
\textbf{\rowVII} &
	\the\numexpr\ENTRYhmmer\relax &
	\the\numexpr\REThmmer\relax &
	\the\numexpr\PATCHhmmer\relax &
	\the\numexpr\TOTALhmmer\relax &
	\the\numexpr\REShmmer\relax &
	\the\numexpr\METAhmmer\relax &
	\the\numexpr\TOTALIIhmmer\relax \\
\textbf{\rowVIII} &
	\the\numexpr\ENTRYsjeng\relax &
	\the\numexpr\RETsjeng\relax &
	\the\numexpr\PATCHsjeng\relax &
	\the\numexpr\TOTALsjeng\relax &
	\the\numexpr\RESsjeng\relax &
	\the\numexpr\METAsjeng\relax &
	\the\numexpr\TOTALIIsjeng\relax \\
\textbf{\rowIX} &
	\the\numexpr\ENTRYlibquantum\relax &
	\the\numexpr\RETlibquantum\relax &
	\the\numexpr\PATCHlibquantum\relax &
	\the\numexpr\TOTALlibquantum\relax &
	\the\numexpr\RESlibquantum\relax &
	\the\numexpr\METAlibquantum\relax &
	\the\numexpr\TOTALIIlibquantum\relax \\
\textbf{\rowX} &
	\the\numexpr\ENTRYhref\relax &
	\the\numexpr\REThref\relax &
	\the\numexpr\PATCHhref\relax &
	\the\numexpr\TOTALhref\relax &
	\the\numexpr\REShref\relax &
	\the\numexpr\METAhref\relax &
	\the\numexpr\TOTALIIhref\relax \\
\textbf{\rowXI} &
	\the\numexpr\ENTRYlbm\relax &
	\the\numexpr\RETlbm\relax &
	\the\numexpr\PATCHlbm\relax &
	\the\numexpr\TOTALlbm\relax &
	\the\numexpr\RESlbm\relax &
	\the\numexpr\METAlbm\relax &
	\the\numexpr\TOTALIIlbm\relax \\
\textbf{\rowXII} &
	\the\numexpr\ENTRYsphinx\relax &
	\the\numexpr\RETsphinx\relax &
	\the\numexpr\PATCHsphinx\relax &
	\the\numexpr\TOTALsphinx\relax &
	\the\numexpr\RESsphinx\relax &
	\the\numexpr\METAsphinx\relax &
	\the\numexpr\TOTALIIsphinx\relax \\
\textbf{\rowXIII} &
	\the\numexpr\ENTRYnginx\relax &
	\the\numexpr\RETnginx\relax &
	\the\numexpr\PATCHnginx\relax &
	\the\numexpr\TOTALnginx\relax &
	\the\numexpr\RESnginx\relax &
	\the\numexpr\METAnginx\relax &
	\the\numexpr\TOTALIInginx\relax \\
\textbf{\rowXIV} &
	\the\numexpr\ENTRYmusl\relax &
	\the\numexpr\RETmusl\relax &
	\the\numexpr\PATCHmusl\relax &
	\the\numexpr\TOTALmusl\relax &
	\the\numexpr\RESmusl\relax &
	\the\numexpr\METAmusl\relax &
	\the\numexpr\TOTALIImusl\relax \\
  \bottomrule
\end{tabular}

  \caption{
    Breakdown of \sys instrumentation
  }
  \spacebelowcaption
  \vspace{-1em}
  \label{tbl:code-stat}
\end{table}
\densetblend

\section{Discussion}
\label{s:discussion}

%%%%%%%%%%%%%%%%%%%%%%%%%%%%%%%%%%%%%%%%%%%%%
\vspace{2px}
\noindent{\bf Applying \sys to binary programs.}
Although \sys requires access to source code,
applying \sys directly to binary programs is possible.
The job of the \sys compiler is
to detect all indirect control transfers (\call/\ret)
and instrument such transfers to utilize trampolines.
Such instrumentation can be done at binary-level
if each \call/\ret can be precisely detected in a program.
Applying \sys to non-PC relative binary is challenging;
however,
position-independent executables (PIEs),
are now the default code generation mode in \cc{gcc-7},
which naturally use PC-relative addressing.
Therefore,
\sys should be practical enough to adopt with little additional effort.

\PP{Full-function reuse attacks}
Throughout our analysis,
we show that existing re-randomization techniques that
use a function trampoline or indirection table,
\ie, use immutable (indirect) code pointer across re-randomization,
cannot prevent full-function reuse attacks.
This also affects \sys;
although limited to functions exposed in the trampoline,
\sys cannot defend against an attacker reusing
such exposed functions as gadgets by leaking code pointers.
We believe that this is a limitation of using immutable code pointers,
and one possible solution to prevent these attacks could be
pairing \sys together with control-flow-integrity (CFI)~\cite{cfi,
zhang:bincfi,
zhang:ccfir,
tice:ifcc,
niu:mcfi,
niu:picfi,
liu:flowguard,
pappas:kbouncer,
van:patharmor,
cheng:ropecker,
prakash:vfguard,
grossklags:tcfi,
gu:pt-cfi,
ge:griffin,
van:typearmor}
or code-pointer integrity/separation (CPI/CPS)~\cite{kuznetsov:cpi}.
\sys already provides backward-edge CFI via shadow stack,
so forward-edge CFI can also be leveraged to further reduce available
code-reuse targets.
\sys's defense is orthogonal to CFI,
so applying both defenses can complement each other
to provide better security.
However, completely eliminating full-function code reuse and
data-oriented programming~\cite{hu-dop} with low performance overhead
and system-wide scalability is still an open problem.
\section{Conclusion}
\label{s:conclusion}
While current defense techniques are capable of warding off current known ROP attacks, most
designs must inherently tradeoff well-rounded performance and scalability
for their security guarantees.
With this insight, we introduce \sys, a novel on-demand system-wide re-randomization
technique to combat code-reuse attacks.
\sys shows pragmatic defense design is indeed navigatable in
the face of the jungle that the code reuse attack landscape is.
\sys is the first code-reuse
defense capable of code-sharing \emph{with} re-randomization and thus allows
scalability in an effort to focus on practicality.
In addition, by being able to re-randomize on-demand, \sys eliminates both the
costly runtime overhead and the integral component of a threshold
associated with continuous re-randomization.
Our evaluation verifies \sys's security guarantees against known
attacker models and adequately quantifies its high-level of entropy.
Furthermore, \sys's performance evaluation showcases its robustness when
deployed with real-world applications derived from SPEC CPU2006 averaging
overhead of 5.5\% as well as confirming scalability in multi-process scenarios
with NGINX web server, averaging only 4.4\% degradation.

%-------------------------------------------------------------------------------
\bibliographystyle{IEEEtranS}
\Urlmuskip=0mu plus 2mu
\bibliography{cfi,sslab,conf}

% Generated by IEEEtranS.bst, version: 1.14 (2015/08/26)
\begin{thebibliography}{10}
\providecommand{\url}[1]{#1}
\csname url@samestyle\endcsname
\providecommand{\newblock}{\relax}
\providecommand{\bibinfo}[2]{#2}
\providecommand{\BIBentrySTDinterwordspacing}{\spaceskip=0pt\relax}
\providecommand{\BIBentryALTinterwordstretchfactor}{4}
\providecommand{\BIBentryALTinterwordspacing}{\spaceskip=\fontdimen2\font plus
\BIBentryALTinterwordstretchfactor\fontdimen3\font minus
  \fontdimen4\font\relax}
\providecommand{\BIBforeignlanguage}[2]{{%
\expandafter\ifx\csname l@#1\endcsname\relax
\typeout{** WARNING: IEEEtranS.bst: No hyphenation pattern has been}%
\typeout{** loaded for the language `#1'. Using the pattern for}%
\typeout{** the default language instead.}%
\else
\language=\csname l@#1\endcsname
\fi
#2}}
\providecommand{\BIBdecl}{\relax}
\BIBdecl

\bibitem{CCS07}
\emph{Proceedings of the 14th ACM Conference on Computer and Communications
  Security (CCS)}, Alexandria, VA, Oct.--Nov. 2007.

\bibitem{SP12}
\emph{Proceedings of the 33rd IEEE Symposium on Security and Privacy
  (Oakland)}, San Francisco, CA, May 2012.

\bibitem{SEC13}
\emph{Proceedings of the 22th USENIX Security Symposium (Security)},
  Washington, DC, Aug. 2013.

\bibitem{SP13}
\emph{Proceedings of the 34th IEEE Symposium on Security and Privacy
  (Oakland)}, San Francisco, CA, May 2013.

\bibitem{SEC14}
\emph{Proceedings of the 23rd USENIX Security Symposium (Security)}, San Diego,
  CA, Aug. 2014.

\bibitem{NDSS15}
\emph{Proceedings of the 2015 Annual Network and Distributed System Security
  Symposium (NDSS)}, San Diego, CA, Feb. 2015.

\bibitem{CCS15}
\emph{Proceedings of the 22nd ACM Conference on Computer and Communications
  Security (CCS)}, Denver, Colorado, Oct. 2015.

\bibitem{SP15}
\emph{Proceedings of the 36th IEEE Symposium on Security and Privacy
  (Oakland)}, San Jose, CA, May 2015.

\bibitem{ASIACCS16}
\emph{Proceedings of the 11th ACM Symposium on Information, Computer and
  Communications Security (ASIACCS)}, Xi'an, China, May--Jun. 2016.

\bibitem{NDSS16}
\emph{Proceedings of the 2016 Annual Network and Distributed System Security
  Symposium (NDSS)}, San Diego, CA, Feb. 2016.

\bibitem{SEC16}
\emph{Proceedings of the 25th USENIX Security Symposium (Security)}, Austin,
  TX, Aug. 2016.

\bibitem{SP16}
\emph{Proceedings of the 37th IEEE Symposium on Security and Privacy
  (Oakland)}, San Jose, CA, May 2016.

\bibitem{musl-web}
``{musl libc},'' 2019, \url{https://wiki.musl-libc.org/}.

\bibitem{SP19}
\emph{Proceedings of the 40th IEEE Symposium on Security and Privacy
  (Oakland)}, San Francisco, CA, May 2019.

\bibitem{cfi}
M.~Abadi, M.~Budiu, U.~Erlingsson, and J.~Ligatti, ``Control-flow integrity,''
  in \emph{Proceedings of the 12th ACM Conference on Computer and
  Communications Security (CCS)}, Alexandria, VA, Nov. 2005.

\bibitem{aleph:stack-smash}
O.~Aleph, ``Smashing the stack for fun and profit,'' \emph{http://www. shmoo.
  com/phrack/Phrack49/p49-14}, 1996.

\bibitem{amazon:c5}
Amazon, ``{Amazon EC2 C5 Instances},'' 2019,
  \url{https://aws.amazon.com/ec2/instance-types/c5/}.

\bibitem{heap-free}
A.~Anonimo, ``Once upon a free ()..'' \emph{Phrack Magazine}, vol.~11, no.~57,
  2001.

\bibitem{arm:XoM}
ARM, ``{ARM Compiler Software Development Guide: 2.21 Execute-only memory},''
  2019,
  \url{http://infocenter.arm.com/help/index.jsp?topic=/com.arm.doc.dui0471j/chr1368698326509.html}.

\bibitem{arm-xom}
{ARM Community}, ``{What is eXecute-Only-Memory (XOM)?}'' july 2017,
  \url{https://community.arm.com/developer/ip-products/processors/b/processors-ip-blog/posts/what-is-execute-only-memory-xom}.

\bibitem{backes:xnr}
M.~Backes, T.~Holz, B.~Kollenda, P.~Koppe, S.~N{\"u}rnberger, and J.~Pewny,
  ``You can run but you can't read: Preventing disclosure exploits in
  executable code,'' in \emph{Proceedings of the 21st ACM Conference on
  Computer and Communications Security (CCS)}, Scottsdale, Arizona, Nov. 2014.

\bibitem{backes:oxymoron}
M.~Backes and S.~N{\"u}rnberger, ``{Oxymoron: Making Fine-Grained Memory
  Randomization Practical by Allowing Code Sharing},'' in \emph{Proceedings of
  the 23rd USENIX Security Symposium (Security)}, San Diego, CA, Aug. 2014.

\bibitem{addrobfuscat-bhatkar-sec05}
S.~Bhatkar, D.~C. DuVarney, and R.~Sekar, ``Efficient techniques for
  comprehensive protection from memory error exploits.'' in \emph{Proceedings
  of the 14th USENIX Security Symposium (Security)}, Baltimore, MD, Aug. 2005.

\bibitem{addressobf-bhatkar-sec03}
------, ``Address obfuscation: An efficient approach to combat a broad range of
  memory error exploits.'' in \emph{Proceedings of the 12th USENIX Security
  Symposium (Security)}, Washington, DC, Aug. 2003.

\bibitem{bigelow:tasr}
D.~Bigelow, T.~Hobson, R.~Rudd, W.~Streilein, and H.~Okhravi, ``{Timely
  Rerandomization for Mitigating Memory Disclosures},'' in \emph{Proceedings of
  the 22nd ACM Conference on Computer and Communications Security (CCS)},
  Denver, Colorado, Oct. 2015.

\bibitem{bittau:brop}
A.~Bittau, A.~Belay, A.~Mashtizadeh, D.~Mazi{\`e}res, and D.~Boneh, ``Hacking
  blind,'' in \emph{Proceedings of the 35th IEEE Symposium on Security and
  Privacy (Oakland)}, San Jose, CA, May 2014.

\bibitem{braden:lr2}
K.~Braden, L.~Davi, C.~Liebchen, A.-R. Sadeghi, S.~Crane, M.~Franz, and
  P.~Larsen, ``{Leakage-Resilient Layout Randomization for Mobile Devices},''
  in \emph{Proceedings of the 2016 Annual Network and Distributed System
  Security Symposium (NDSS)}, San Diego, CA, Feb. 2016.

\bibitem{shadesmar-burow-oakland19}
N.~Burow, X.~Zhang, and M.~Payer, ``{SoK: Shining Light on Shadow Stacks},'' in
  \emph{Proceedings of the 40th IEEE Symposium on Security and Privacy
  (Oakland)}, San Francisco, CA, May 2019.

\bibitem{chen:codearmor}
X.~Chen, H.~Bos, and C.~Giuffrida, ``{CodeArmor: Virtualizing The Code Space to
  Counter Disclosure Attacks},'' in \emph{Proceedings of the 2nd IEEE European
  Symposium on Security and Privacy (Euro S\&P)}, Paris, France, Apr. 2017.

\bibitem{cheng:ropecker}
Y.~Cheng, Z.~Zhou, Y.~Miao, X.~Ding, and R.~H. Deng, ``Ropecker: A generic and
  practical approach for defending against rop attack,'' Feb. 2014.

\bibitem{conti:selfrando}
M.~Conti, S.~Crane, T.~Frassetto, A.~Homescu, G.~Koppen, P.~Larsen,
  C.~Liebchen, M.~Perry, and A.-R. Sadeghi, ``Selfrando: Securing the tor
  browser against de-anonymization exploits,'' \emph{Proceedings on Privacy
  Enhancing Technologies}, vol. 2016, no.~4, pp. 454--469, 2016.

\bibitem{crane:readactor}
S.~Crane, C.~Liebchen, A.~Homescu, L.~Davi, P.~Larsen, A.-R. Sadeghi,
  S.~Brunthaler, and M.~Franz, ``{Readactor: Practical Code Randomization
  Resilient to Memory Disclosure},'' in \emph{Proceedings of the 36th IEEE
  Symposium on Security and Privacy (Oakland)}, San Jose, CA, May 2015.

\bibitem{crane:readactor++}
S.~J. Crane, S.~Volckaert, F.~Schuster, C.~Liebchen, P.~Larsen, L.~Davi, A.-R.
  Sadeghi, T.~Holz, B.~De~Sutter, and M.~Franz, ``{It's a TRaP: Table
  Randomization and Protection Against Function-reuse Attacks},'' in
  \emph{Proceedings of the 36th IEEE Symposium on Security and Privacy
  (Oakland)}, San Jose, CA, May 2015.

\bibitem{davi:isomeron}
L.~Davi, C.~Liebchen, A.-R. Sadeghi, K.~Z. Snow, and F.~Monrose, ``{Isomeron:
  Code Randomization Resilient to (Just-In-Time) Return-Oriented
  Programming},'' in \emph{Proceedings of the 2015 Annual Network and
  Distributed System Security Symposium (NDSS)}, San Diego, CA, Feb. 2015.

\bibitem{evans:missingcpi}
I.~Evans, S.~Fingeret, J.~Gonzalez, U.~Otgonbaatar, T.~Tang, H.~Shrobe,
  S.~Sidiroglou-Douskos, M.~Rinard, and H.~Okhravi, ``Missing the point {(er)}:
  On the effectiveness of code pointer integrity,'' in \emph{Proceedings of the
  36th IEEE Symposium on Security and Privacy (Oakland)}, San Jose, CA, May
  2015.

\bibitem{hardening-fedora}
{Fedora}, ``{Hardening Flags Updates for Fedora 28},'' 2018,
  \url{https://fedoraproject.org/wiki/Changes/HardeningFlags28}.

\bibitem{morpheus-gallagher-asplos19}
M.~Gallagher, L.~Biernacki, S.~Chen, Z.~B. Aweke, S.~F. Yitbarek, M.~T. Aga,
  A.~Harris, Z.~Xu, B.~Kasikci, V.~Bertacco, S.~Malik, M.~Tiwari, and
  T.~Austin, ``{Morpheus: A Vulnerability-Tolerant Secure Architecture Based on
  Ensembles of Moving Target Defenses with Churn},'' in \emph{Proceedings of
  the 24th ACM International Conference on Architectural Support for
  Programming Languages and Operating Systems (ASPLOS)}, Providence, RI, USA,
  Apr. 2019, pp. 469--484.

\bibitem{gawlik:crop}
R.~Gawlik, B.~Kollenda, P.~Koppe, B.~Garmany, and T.~Holz, ``Enabling
  client-side crash-resistance to overcome diversification and information
  hiding.'' in \emph{Proceedings of the 2016 Annual Network and Distributed
  System Security Symposium (NDSS)}, San Diego, CA, Feb. 2016.

\bibitem{gawlik:crash-resist}
------, ``{Enabling Client-Side Crash-Resistance to Overcome Diversification
  and Information Hiding},'' in \emph{Proceedings of the 2016 Annual Network
  and Distributed System Security Symposium (NDSS)}, San Diego, CA, Feb. 2016.

\bibitem{ge:griffin}
X.~Ge, W.~Cui, and T.~Jaeger, ``Griffin: Guarding control flows using intel
  processor trace,'' in \emph{Proceedings of the 22nd ACM International
  Conference on Architectural Support for Programming Languages and Operating
  Systems (ASPLOS)}, Xi'an, China, Apr. 2017.

\bibitem{gionta:hidem}
J.~Gionta, W.~Enck, and P.~Ning, ``{HideM: Protecting the Contents of Userspace
  Memory in the Face of Disclosure Vulnerabilities},'' in \emph{Proceedings of
  the 5th ACM Conference on Data and Application Security and Privacy
  (CODASPY)}, San Antonio, TX, Mar. 2015.

\bibitem{giuffrida:asr3}
C.~Giuffrida, A.~Kuijsten, and A.~S. Tanenbaum, ``{Enhanced Operating System
  Security Through Efficient and Fine-grained Address Space Randomization},''
  in \emph{Proceedings of the 21st USENIX Security Symposium (Security)},
  Bellevue, WA, Aug. 2012.

\bibitem{github:wrk}
W.~Glozer, ``{a HTTP benchmarking tool},'' 2019,
  \url{https://github.com/wg/wrk}.

\bibitem{goktas:information-hiding}
E.~G{\"o}kta{\c s}, R.~Gawlik, B.~Kollenda, E.~Athanasopoulos, G.~Portokalidis,
  C.~Giuffrida, and H.~Bos, ``{Undermining Information Hiding (and What to Do
  about It)},'' in \emph{Proceedings of the 25th USENIX Security Symposium
  (Security)}, Austin, TX, Aug. 2016.

\bibitem{grossklags:tcfi}
J.~Grossklags and C.~Eckert, ``$\tau${CFI: Type-Assisted Control Flow Integrity
  for x86-64 Binaries},'' in \emph{Proceedings of the 21th International
  Symposium on Research in Attacks, Intrusions and Defenses (RAID)}, Heraklion,
  Crete, Greece, Sep. 2018.

\bibitem{gu:pt-cfi}
Y.~Gu, Q.~Zhao, Y.~Zhang, and Z.~Lin, ``Pt-cfi: Transparent backward-edge
  control flow violation detection using intel processor trace,'' in
  \emph{Proceedings of the 7th ACM Conference on Data and Application Security
  and Privacy (CODASPY)}, Scottsdale, AZ, Mar. 2017.

\bibitem{hiser:ilr}
J.~Hiser, A.~Nguyen-Tuong, M.~Co, M.~Hall, and J.~W. Davidson, ``{ILR: Where'd
  My Gadgets Go?}'' in \emph{Proceedings of the 33rd IEEE Symposium on Security
  and Privacy (Oakland)}, San Francisco, CA, May 2012.

\bibitem{librando-homescu-ccs13}
A.~Homescu, S.~Brunthaler, P.~Larsen, and M.~Franz, ``{librando: Transparent
  Code Randomization for Just-in-Time Compilers},'' in \emph{Proceedings of the
  20th ACM Conference on Computer and Communications Security (CCS)}, Berlin,
  Germany, Oct. 2013, pp. 993--1004.

\bibitem{hu-dop}
H.~Hu, S.~Shinde, S.~Adrian, Z.~L. Chua, P.~Saxena, and Z.~Liang,
  ``Data-oriented programming: On the expressiveness of non-control data
  attacks,'' in \emph{Security and Privacy (SP), 2016 IEEE Symposium on}.\hskip
  1em plus 0.5em minus 0.4em\relax IEEE, 2016, pp. 969--986.

\bibitem{intel:sdm}
{Intel Corporation}, ``{Intel 64 and IA-32 Architectures Software Developer's
  Manual},'' 2019, \url{https://software.intel.com/en-us/articles/intel-sdm}.

\bibitem{intel:xeon-scalable}
------, ``{INTEL~\textregistered~XEON~\textregistered~SCALABLE PROCESSORS},''
  2019,
  \url{https://www.intel.com/content/www/us/en/products/processors/xeon/scalable.html}.

\bibitem{lwn:dep}
{Jonathan Corbet}, ``{x86 NX support},'' 2004,
  \url{https://lwn.net/Articles/87814/}.

\bibitem{kaempf:heap-vudo}
M.~Kaempf, ``Vudo malloc tricks. phrack magazine, 57 (8), august 2001.''

\bibitem{kil:aslp}
C.~Kil, J.~Jun, C.~Bookholt, J.~Xu, and P.~Ning, ``Address space layout
  permutation ({ASLP}): Towards fine-grained randomization of commodity
  software,'' in \emph{Proceedings of the Annual Computer Security Applications
  Conference (ACSAC)}, 2006.

\bibitem{spectre-kocher-sp19}
P.~Kocher, J.~Horn, A.~Fogh, , D.~Genkin, D.~Gruss, W.~Haas, M.~Hamburg,
  M.~Lipp, S.~Mangard, T.~Prescher, M.~Schwarz, and Y.~Yarom, ``Spectre
  attacks: Exploiting speculative execution,'' in \emph{Proceedings of the 40th
  IEEE Symposium on Security and Privacy (Oakland)}, San Francisco, CA, May
  2019.

\bibitem{kollenda:auto-crash-resist}
B.~Kollenda, E.~G{\"o}kta{\c{s}}, T.~Blazytko, P.~Koppe, R.~Gawlik, R.~K.
  Konoth, C.~Giuffrida, H.~Bos, and T.~Holz, ``{Towards Automated Discovery of
  Crash-resistant Primitives in Binary Executables},'' in \emph{Proceedings of
  the 47th International Conference on Dependable Systems and Networks (DSN)},
  Denver, CO, Jun. 2017.

\bibitem{koo:juggling}
H.~Koo and M.~Polychronakis, ``Juggling the gadgets: Binary-level code
  randomization using instruction displacement,'' in \emph{Proceedings of the
  11th ACM Symposium on Information, Computer and Communications Security
  (ASIACCS)}, Xi'an, China, May--Jun. 2016.

\bibitem{kuznetsov:cpi}
V.~Kuznetsov, L.~Szekeres, M.~Payer, G.~Candea, R.~Sekar, and D.~Song,
  ``{Code-Pointer Integrity},'' in \emph{Proceedings of the 11th USENIX
  Symposium on Operating Systems Design and Implementation (OSDI)}, Broomfield,
  Colorado, Oct. 2014.

\bibitem{meltdown-lipp-sec18}
M.~Lipp, M.~Schwarz, D.~Gruss, T.~Prescher, W.~Haas, A.~Fogh, J.~Horn,
  S.~Mangard, P.~Kocher, D.~Genkin, Y.~Yarom, and M.~Hamburg, ``Meltdown:
  Reading kernel memory from user space,'' in \emph{Proceedings of the 27th
  USENIX Security Symposium (Security)}, Baltimore, MD, Aug. 2018.

\bibitem{liu:flowguard}
Y.~Liu, P.~Shi, X.~Wang, H.~Chen, B.~Zang, and H.~Guan, ``Transparent and
  efficient {CFI} enforcement with intel processor trace,'' in
  \emph{Proceedings of the 23rd IEEE Symposium on High Performance Computer
  Architecture (HPCA)}, Austin, TX, Feb. 2017.

\bibitem{lu:runtimeASLR}
K.~Lu, W.~Lee, S.~N{\"u}rnberger, and M.~Backes, ``{How to Make ASLR Win the
  Clone Wars: Runtime Re-Randomization},'' in \emph{Proceedings of the 2016
  Annual Network and Distributed System Security Symposium (NDSS)}, San Diego,
  CA, Feb. 2016.

\bibitem{mpk-glibc-larabel}
{Michael Larabel}, ``{Glibc Rolls Out Support For Memory Protection Keys},''
  2017,
  \url{https://www.phoronix.com/scan.php?page=news_item&px=Glibc-Memory-Protection-Keys}.

\bibitem{windows:dep}
{Microsoft Support}, ``{A detailed description of the Data Execution Prevention
  (DEP) feature in Windows XP Service Pack 2, Windows XP Tablet PC Edition
  2005, and Windows Server 2003},'' 2017,
  \url{https://support.microsoft.com/en-us/help/875352/a-detailed-description-of-the-data-execution-prevention-dep-feature-in}.

\bibitem{niu:mcfi}
B.~Niu and G.~Tan, ``Modular control-flow integrity,'' in \emph{Proceedings of
  the 2014 ACM SIGPLAN Conference on Programming Language Design and
  Implementation (PLDI)}, Edinburgh, UK, Jun. 2014.

\bibitem{niu:picfi}
------, ``Per-input control-flow integrity,'' in \emph{Proceedings of the 22nd
  ACM Conference on Computer and Communications Security (CCS)}, Denver,
  Colorado, Oct. 2015.

\bibitem{oikonomopoulos:pokingcpi}
A.~Oikonomopoulos, E.~Athanasopoulos, H.~Bos, and C.~Giuffrida, ``Poking holes
  in information hiding.'' in \emph{Proceedings of the 25th USENIX Security
  Symposium (Security)}, Austin, TX, Aug. 2016.

\bibitem{pappas:smashing}
V.~Pappas, M.~Polychronakis, and A.~D. Keromytis, ``Smashing the gadgets:
  Hindering return-oriented programming using in-place code randomization,'' in
  \emph{Proceedings of the 33rd IEEE Symposium on Security and Privacy
  (Oakland)}, San Francisco, CA, May 2012.

\bibitem{pappas:kbouncer}
------, ``{Transparent ROP Exploit Mitigation Using Indirect Branch Tracing},''
  in \emph{Proceedings of the 22th USENIX Security Symposium (Security)},
  Washington, DC, Aug. 2013.

\bibitem{bgdx-pewny-acsac17}
J.~Pewny, P.~Koppe, L.~Davi, and T.~Holz, ``{Breaking and Fixing Destructive
  Code Read Defenses},'' in \emph{Proceedings of the 12th ACM Symposium on
  Information, Computer and Communications Security (ASIACCS)}, Abu Dhabi, UAE,
  Apr. 2017, pp. 55--67.

\bibitem{pomonis:krx}
M.~Pomonis, T.~Petsios, A.~D. Keromytis, M.~Polychronakis, and V.~P. Kemerlis,
  ``{kR\^{} X: Comprehensive Kernel Protection against Just-In-Time Code
  Reuse},'' in \emph{Proceedings of the 12th European Conference on Computer
  Systems (EuroSys)}, Belgrade, Serbia, Apr. 2017.

\bibitem{prakash:vfguard}
A.~Prakash, X.~Hu, and H.~Yin, ``{vfGuard: Strict Protection for Virtual
  Function Calls in COTS C++ Binaries.}'' in \emph{Proceedings of the 2015
  Annual Network and Distributed System Security Symposium (NDSS)}, San Diego,
  CA, Feb. 2015.

\bibitem{rudd:aocr}
R.~Rudd, R.~Skowyra, D.~Bigelow, V.~Dedhia, T.~Hobson, S.~Crane, C.~Liebchen,
  P.~Larsen, L.~Davi, M.~Franz \emph{et~al.}, ``{Address-Oblivious Code Reuse:
  On the Effectiveness of Leakage Resilient Diversity},'' in \emph{Proceedings
  of the 2017 Annual Network and Distributed System Security Symposium (NDSS)},
  San Diego, CA, Feb.--Mar. 2017.

\bibitem{ROPGadget}
J.~Salwan, ``Ropgadget: Gadgets finder and auto-roper,'' 2019,
  \url{https://github.com/JonathanSalwan/ROPgadget}.

\bibitem{shacham:rop}
H.~Shacham, ``The geometry of innocent flesh on the bone: Return-into-libc
  without function calls (on the x86),'' in \emph{Proceedings of the 14th ACM
  Conference on Computer and Communications Security (CCS)}, Alexandria, VA,
  Oct.--Nov. 2007.

\bibitem{shacham:rintolibc}
------, ``The geometry of innocent flesh on the bone: Return-into-libc without
  function calls (on the x86),'' in \emph{Proceedings of the 14th ACM
  Conference on Computer and Communications Security}, Alexandria, VA,
  Oct.--Nov. 2007.

\bibitem{zombie-gadget-snow-oakland16}
K.~Z. {Snow}, R.~{Rogowski}, J.~{Werner}, H.~{Koo}, F.~{Monrose}, and
  M.~{Polychronakis}, ``{Return to the Zombie Gadgets: Undermining Destructive
  Code Reads via Code Inference Attacks},'' in \emph{Proceedings of the 37th
  IEEE Symposium on Security and Privacy (Oakland)}, San Jose, CA, May 2016.

\bibitem{snow:jitrop}
K.~Z. Snow, F.~Monrose, L.~Davi, A.~Dmitrienko, C.~Liebchen, and A.-R. Sadeghi,
  ``{Just-in-time Code Reuse: On the Effectiveness of Fine-grained Address
  Space Layout Randomization},'' in \emph{Proceedings of the 34th IEEE
  Symposium on Security and Privacy (Oakland)}, San Francisco, CA, May 2013.

\bibitem{tang:heisenbyte}
A.~Tang, S.~Sethumadhavan, and S.~Stolfo, ``Heisenbyte: Thwarting memory
  disclosure attacks using destructive code reads,'' in \emph{Proceedings of
  the 22nd ACM Conference on Computer and Communications Security (CCS)},
  Denver, Colorado, Oct. 2015.

\bibitem{pax-aslr}
{The PAX Team}, ``{Address Space Layout Randomization},'' 2003,
  \url{https://pax.grsecurity.net/docs/aslr.txt}.

\bibitem{tice:ifcc}
C.~Tice, T.~Roeder, P.~Collingbourne, S.~Checkoway, {\'U}.~Erlingsson,
  L.~Lozano, and G.~Pike, ``{Enforcing Forward-Edge Control-Flow Integrity in
  GCC \& LLVM},'' in \emph{Proceedings of the 23rd USENIX Security Symposium
  (Security)}, San Diego, CA, Aug. 2014.

\bibitem{van:patharmor}
V.~van~der Veen, D.~Andriesse, E.~G{\"o}kta{\c{s}}, B.~Gras, L.~Sambuc,
  A.~Slowinska, H.~Bos, and C.~Giuffrida, ``Practical context-sensitive
  {CFI},'' in \emph{Proceedings of the 22nd ACM Conference on Computer and
  Communications Security (CCS)}, Denver, Colorado, Oct. 2015.

\bibitem{van:newton}
V.~van~der Veen, D.~Andriesse, M.~Stamatogiannakis, X.~Chen, H.~Bos, and
  C.~Giuffrdia, ``The dynamics of innocent flesh on the bone: Code reuse ten
  years later,'' in \emph{Proceedings of the 24th ACM Conference on Computer
  and Communications Security (CCS)}, Dallas, TX, Oct.--Nov. 2017.

\bibitem{van:typearmor}
V.~van~der Veen, E.~G{\"o}ktas, M.~Contag, A.~Pawoloski, X.~Chen, S.~Rawat,
  H.~Bos, T.~Holz, E.~Athanasopoulos, and C.~Giuffrida, ``A tough call:
  Mitigating advanced code-reuse attacks at the binary level,'' in
  \emph{Proceedings of the 37th IEEE Symposium on Security and Privacy
  (Oakland)}, San Jose, CA, May 2016.

\bibitem{wang:reranz}
Z.~Wang, C.~Wu, J.~Li, Y.~Lai, X.~Zhang, W.-C. Hsu, and Y.~Cheng, ``{Reranz: A
  Light-weight Virtual Machine to Mitigate Memory Disclosure Attacks},'' in
  \emph{Proceedings of the 13th International Conference on Virtual Execution
  Environments (VEE)}, Xi'an, China, Apr. 2017.

\bibitem{relrop-wardleakage-esorics19}
B.~C. Ward, R.~Skowyra, C.~Spensky, J.~Martin, and H.~Okhravi, ``{The
  Leakage-Resilience Dilemma},'' in \emph{Proceedings of the 24th European
  Symposium on Research in Computer Security (ESORICS)}, Luxembourg, Sep. 2019.

\bibitem{wartell:stirring}
R.~Wartell, V.~Mohan, K.~W. Hamlen, and Z.~Lin, ``Binary stirring:
  Self-randomizing instruction addresses of legacy x86 binary code,'' in
  \emph{Proceedings of the 19th ACM Conference on Computer and Communications
  Security (CCS)}, Raleigh, NC, Oct. 2012.

\bibitem{werner:near}
J.~Werner, G.~Baltas, R.~Dallara, N.~Otterness, K.~Z. Snow, F.~Monrose, and
  M.~Polychronakis, ``No-execute-after-read: Preventing code disclosure in
  commodity software,'' in \emph{Proceedings of the 11th ACM Symposium on
  Information, Computer and Communications Security (ASIACCS)}, Xi'an, China,
  May--Jun. 2016.

\bibitem{williams:shuffler}
D.~Williams-King, G.~Gobieski, K.~Williams-King, J.~P. Blake, X.~Yuan, P.~Colp,
  M.~Zheng, V.~P. Kemerlis, J.~Yang, and W.~Aiello, ``{Shuffler: Fast and
  Deployable Continuous Code Re-Randomization},'' in \emph{Proceedings of the
  12th USENIX Symposium on Operating Systems Design and Implementation (OSDI)},
  Savannah, GA, Nov. 2016.

\bibitem{zhang:ccfir}
C.~Zhang, T.~Wei, Z.~Chen, L.~Duan, L.~Szekeres, S.~McCamant, D.~Song, and
  W.~Zou, ``Practical control flow integrity and randomization for binary
  executables,'' in \emph{Proceedings of the 34th IEEE Symposium on Security
  and Privacy (Oakland)}, San Francisco, CA, May 2013.

\bibitem{zhang:bincfi}
M.~Zhang and R.~Sekar, ``{Control Flow Integrity for COTS Binaries},'' in
  \emph{Proceedings of the 22th USENIX Security Symposium (Security)},
  Washington, DC, Aug. 2013.

\end{thebibliography}
%%%%%%%%%%%%%%%%%%%%%%%%%%%%%%%%%%%%%%%%%%%%%%%%%%%%%%%%%%%%%%%%%%%%%%%%%%%%%%%%
\end{document}